\documentclass[iop]{emulateapj}

\usepackage{times,psfig,epsfig} 
\usepackage{rotating, graphicx,dsfont}
\usepackage{color}
\usepackage{amssymb, amsmath}

\usepackage[utf8]{inputenc}
\usepackage{lipsum}
\usepackage{pifont}
\usepackage{amsmath, amssymb, bm}
\usepackage{balance}
\usepackage{multirow}
\usepackage{enumitem}
\usepackage{booktabs}

\shorttitle{The imprint of neutrinos on clustering in redshift-space}
\shortauthors{Francisco Villaescusa-Navarro et al.}

\usepackage[usenames,dvipsnames]{xcolor}
\usepackage{hyperref}
\hypersetup{
    colorlinks = true,
    citecolor = {MidnightBlue},
    linkcolor = {BrickRed},
    urlcolor = {black}		
}

\newcommand{\be}{\begin{equation}}
\newcommand{\ee}{\end{equation}}
\newcommand{\ba}{\begin{eqnarray}}
\newcommand{\ea}{\end{eqnarray}}

\begin{document}

\title{The imprint of neutrinos on clustering in redshift-space}

\author{Francisco Villaescusa-Navarro$^{1,\dagger}$, Arka Banerjee$^{2}$, Neal Dalal$^{2}$, Emanuele Castorina$^3$, \\Roman Scoccimarro$^4$, Raul Angulo$^5$ and David N. Spergel$^{6,1}$\\}
\affil{$^{1}${Center for Computational Astrophysics, Flatiron Institute, 162 5th Avenue, 10010, New York, NY, USA}}
\affil{$^{2}${Department of Physics, University of Illinois at Urbana-Champaign, 1110 West Green Street, Urbana, IL 61801-3080 USA}}
\affil{$^{3}${Berkeley Center for Cosmological Physics Campbell Hall 341, University of California, Berkeley CA 94720}}
\affil{$^{4}${Center for Cosmology and Particle Physics, Department of Physics, New York University, NY 10003, New York, USA}}
\affil{$^{5}${Centro de Estudios de F\'isica del Cosmos de Arag\'on, Plaza San Juan 1, Planta-2, 44001, Teruel, Spain}}
\affil{$^{6}${Department of Astrophysical Sciences, Princeton University, Peyton Hall, Princeton NJ 08544-0010, USA}}
\altaffiltext{$\dagger$}{fvillaescusa@flatironinstitute.org}

\begin{abstract}
We investigate the signatures left by the cosmic neutrino background on the clustering of matter, CDM+baryons and halos in redshift-space using the \textsc{HADES} simulations: a set of more than 1000 N-body and hydrodynamical simulations with massless and massive neutrinos. While on large scales the clustering of matter and CDM+baryons is very different in cosmologies with massive and massless neutrinos, we find that the effect neutrinos induce on the clustering of CDM+baryons in redshift-space on small scales is almost entirely due to the change in $\sigma_8$.  However, neutrinos do imprint a characteristic signature in the quadrupole of the {\em total} matter field (CDM+baryon+neutrinos) on small scales, that can be used to disentangle the effect of $\sigma_8$ and $M_\nu$. We show that the effect of neutrinos on the clustering of halos is very different, on all scales, to the effects induced by varying $\sigma_8$. We find that the effects of neutrinos of the growth rate of CDM+baryons ranges from $\sim0.3\%$ to $2\%$ on scales $k\in[0.01, 0.5]~h{\rm Mpc}^{-1}$ for neutrinos with masses $M_\nu \leqslant 0.15$ eV. We compute the bias between the momentum of halos and the momentum of CDM+baryon and find it to be 1 on large scales for all models with massless and massive neutrinos considered. This point towards a velocity bias between halos and total matter on large scales that it is important to account for in order to extract unbiased neutrino information from velocity/momentum surveys such as kSZ observations. We show that, even on very large-scales, non-linear corrections are important to describe the clustering of halos in redshift-space in cosmologies with massless and massive neutrinos at low redshift. 

We show that baryonic effects can affect the clustering of matter and CDM+baryons in redshift-space by up to a few percent down to $k=0.5~h{\rm Mpc}^{-1}$. We find that hydrodynamics and astrophysical processes, as implemented in our simulations, only distort the relative effect that neutrinos induce on the anisotropic clustering of matter, CDM+baryons and halos in redshift-space by less than $1\%$. Thus, the effect of neutrinos in the fully non-linear regime can be written as a transfer function with very weak dependence on astrophysics that can be studied through N-body simulations.
\end{abstract} 

\keywords{neutrinos -- cosmology: cosmological parameters -- methods: numerical}

\section{Introduction}
\label{sec:introduction}

The Standard Model of particle physics describes neutrinos as fundamental massless particles. Neutrino oscillation experiments have shown, however, that at least two neutrino families have mass, with a lower limit on the sum of the neutrino masses \citep{Concha_2014,Forero_2014, Esteban_2017}: $M_\nu=\sum_i m_{\nu_i}\gtrsim0.06$ eV. Unfortunately, oscillation measurements are not sensitive to the absolute mass scale of the neutrinos, requiring the use of alternative probes. Tritium beta decay experiments can be used to place upper limits on neutrino masses, with current constraints at the level $M_\nu\lesssim6.9$ eV \citep{Kraus_2005}.

Neutrinos are among the most abundant particles in the Universe, with number densities only somewhat less than photons.  Unlike photons, however, at least some neutrinos have rest mass, implying that relic neutrinos can produce significant effects on cosmological observables, in particular the low-redshift evolution of cosmological density perturbations.  One key difference between neutrinos and other gravitating massive species, like baryons or cold dark matter (CDM), is that neutrinos have large thermal velocities, which can lead to distinctive signatures of neutrinos in many different cosmological observables. Those signatures have been used to place tight upper limits on the sum of the neutrino masses. The current tightest bounds have been obtained by combining different cosmological observables such as the anisotropies in the cosmic microwave background, galaxy clustering and the Ly$\alpha$ forest: $M_\nu\lesssim0.12$ eV \citep{Palanque_2015, Cuesta_2016, Vagnozzi_2017}, significantly tighter than the bounds possible from terrestrial experiments for the foreseeable future.
 
These constraints, as impressive as they are, will only improve in the near future as additional cosmological surveys come online.  In particular, surveys like \textsc{Euclid}\footnote{http://sci.esa.int/euclid/}, \textsc{DESI}\footnote{http://desi.lbl.gov/}, \textsc{wfirst}\footnote{https://wfirst.gsfc.nasa.gov/} and \textsc{PFS}\footnote{http://sumire.ipmu.jp/en/2652/} will measure the 3-dimensional clustering of galaxies across a range of length scales and cosmic time.  As is well known, 3-D clustering represents one of the most important sources of cosmological information. Besides constraining neutrino masses, galaxy clustering also provides a wealth ofinformation about many crucial cosmological questions, including the energy content of the Universe, the initial conditions laid down during inflation, and the overall spatial geometry of the universe.  
However, multiple physical processes can impede both the measurement of  matter clustering and its interpretation. Firstly, at late times and on small scales, matter clustering becomes a highly nonlinear process, and may be contaminated by poorly-characterized astrophysical processes such as feedback from star formation, supernovae, and AGN.  Secondly, the dominant matter component, CDM, is not directly visible, meaning that the underlying clustering of matter typically must be indirectly inferred through biased tracers such as galaxies or neutral hydrogen. Thirdly, peculiar velocities induce a shift in the redshifts we measure, distorting the clustering pattern and breaking statistical isotropy of matter 2-point clustering.

In spite of these difficulties, the cosmolgical surveys mentioned above will cover volumes so large that they will be able to either measure the neutrino masses or to place extremely tight constraints on them \citep[e.g.][]{Audren_2013, McDonald_2014, Villaescusa-Navarro_2015, Petracca_2016, Allison_2015, Sartoris_2016}.  Given the statistical power of the upcoming generation of surveys, it is becoming increasingly crucial to have accurate theoretical predictions \citep{Tobias_2016} that allow us to understand and to model accurately the impact of non-linearities, galaxy bias and redshift-space distortions in cosmologies with massive neutrinos.
The impact of neutrino masses on the fully non-linear clustering of matter in real-space has been carefully studied in a number of different works \citep{Saito_2008, Brandbyge_2008, Wong_2008, Saito_2009, Brandbyge_2010, Rossi_2014, LoVerde_2014, Agarwal2011,Bird_2011,Wagner2012,Viel_2010,Villaescusa-Navarro_2012, Yacine-Bird, LesgourguesBook, Blas_2014, Inman_2015, Wong_2015, Peloso_2015, Viel_2010, Rossi_2014, Upadhye_2014, Castorina_2015, Villaescusa-Navarro_2015, Arka_2016, Emberson_2016} and the halo model \citep{Cooray_Sheth_2002} has been extended to cosmologies with massive neutrinos \citep{Massara_2014}. 
In particular,
halo bias in cosmologies with massive neutrinos has been recently studied in detail; it has been shown that neutrino masses induce a scale-dependent bias on large-scales \citep{Villaescusa-Navarro_2014, Castorina_2014, LoVerde_2014b, Biagetti_2014, Castorina_2015}. This happens because the transfer functions of neutrinos and CDM+baryons are different in shape and amplitude, even on large scales. Since halo properties are dictated by the statistics of underlying density field of CDM+baryons, and not by the total matter field (which includes neutrinos) \citep{Ichiki-Takada, Castorina_2014, LoVerde_2014}, halo bias is sensitive to whether it is defined with respect to the CDM+baryon field or the total matter field. Not modeling properly this effect can lead to biases in the estimated values of the cosmological parameters and neutrino masses \citep{Raccanelli_2017}. 
\cite{Arka_2016} have pointed out that neutrinos have a larger effect on the clustering of cosmic voids than in halos; they induce a stronger scale-dependent bias on large scales. The reason behind is that neutrinos affect more intensively voids than halos \citep{Massara_2015}, since they barely cluster in the latter but can not evacuate the former.

The purpose of this work is to investigate whether neutrinos imprint characteristic signatures on the clustering of matter, CDM+baryons and halos, in redshift-space. We note that previous works have studied redshift-space distortions in cosmologies with massive neutrinos \citep{Marulli_2011, Upadhye_2016, Castorina_2015}.  The goal of this work, in relation to previous investigations, is a more systematic study is carried out using a much larger set of simulations covering large volumes. We also study for the first time the impact of neutrinos on the growth rate of cosmic structure, the extent of velocity or momentum bias generated by neutrinos, and the effect of baryons on inference of neutrino properties.

We employ a set of more than 1000 state-of-the-art N-body and hydrodynamic simulations, with realistic neutrino masses, covering volumes larger than $100~(h^{-1}{\rm Gpc})^3$. We try to answer key questions such as the information content on neutrino masses embedded into galaxy clustering in redshift-space. An important limitation to that information content is the presence of degeneracies, among which the $M_\nu$-$\sigma_8$ is one of the most prominents\footnote{It is well known that the $M_\nu$-$\sigma_8$ degeneracy can be broken by adding information from other sources, such as CMB data \cite[e.g.][]{Peloso_2015}}. 

We focus our analysis on the effect induced by neutrinos on the monopole, quadrupole and fully 2D power spectrum of matter, CDM+baryon and halo fields. We also estimate the amplitude and shape of the growth rate of the CDM+baryon field, an important ingredient to model the galaxy power spectrum in redshift-space. 

We compute the bias between the momentum of halos and the CDM+baryon fields and find that there is no detectable large-scale velocity bias between those fields generated by massive neutrinos. This implies that a bias between the momentum of halos the total matter field will be present in models with massive neutrinos. Finally, we study the range of validity of linear theory to model the clustering of halos in redshift-space in models with massive neutrinos. 

We also study the impact of baryonic processes on redshift-space distortions in models with massive neutrinos. We investigate the impact of baryonic processes on both the absolute and relative amplitude and shape of the clustering pattern of the matter, CDM+baryons and halo fields in redshift-space.

This paper is organized as follows. In section \ref{sec:simulations} we describe the simulation suite we have used in this paper. The relative differences induced by the neutrino masses considered in this paper are much smaller than the overall amplitude of the quantities we consider. Thus, in this paper we focus in relative differences and we show absolute quantities in section \ref{sec:features}. In section \ref{sec:matter} we investigate the impact of neutrino masses of the matter and CDM+baryon density fields and estimate the growth rate of CDM+baryons in the fully non-linear regime. The effects of neutrinos on the clustering of halos, the momentum bias and the validity of linear theory are shown in section \ref{sec:halos}. In section \ref{sec:baryons} we study the impact of baryonic effects on the distribution of matter, CDM+baryons and halos in redshift-space. We draw the main conclusions of this work in section \ref{sec:conclusions}.

\section{Numerical simulations}
\label{sec:simulations}

\begin{table*}
\begin{center}
\caption{\label{tab:i} Specifications of the simulation suite used. All simulations share the value of the following cosmological parameters: $\Omega_{\rm m}=\Omega_{\rm c}+\Omega_{\rm b}+\Omega_\nu=0.3175$, $\Omega_{\rm b}=0.049$, $\Omega_\Lambda=0.6825$, $n_s=0.9624$, $h=0.6711$. The values of $M_\nu$, $A_s$ and $\sigma_8$ for each model are given in the columns 2-5. The simulations follow the evolution of $N_{\rm cdm}$ CDM particles, $N_{\rm gas}$ gas particles and $N_\nu$ neutrino particles, with masses $m_{\rm cdm}$, $m_{\rm gas}$ and $m_\nu$, within a box size of 1 $h^{-1}{\rm Gpc}$. The softening length for both CDM and neutrinos is given by $\epsilon$, while for gas is set to the SPH radius. The number of realizations for each model is shown in the last column.  Simulations without gas particles are N-body while those with gas particles are hydrodynamic.}
{\renewcommand{\arraystretch}{1.3}
\begin{tabular}{| c | c | c | c | c | c | c | c | c | c | c | c | c |}
\hline
Name & $\sum m_\nu$ & $10^9A_s$ & $\sigma_8^m$ & $\sigma_8^c$ & $N_{\rm cdm}^{1/3}$ & $N_{\rm gas}^{1/3}$ & $N_\nu^{1/3}$  & $m_{\rm cdm}$ & $m_{\rm gas}$ & $m_\nu$ & $\epsilon$ & realizations\\
&  (eV) & & & & & & & $(10^{10}h^{-1}M_\odot)$ & $(10^{10}h^{-1}M_\odot)$ & $(10^8h^{-1}M_\odot)$ & $(h^{-1}{\rm kpc})$ & \\
\hline \hline

\multirow{3}{*}{L0 (fid)} & \multirow{3}{*}{0.00} & \multirow{3}{*}{2.13} & \multicolumn{2}{|c|}{\multirow{3}{*}{0.833}} & $512$ & 0 & $0$ & $65.66$ & 0 &  0 &50 & 100\\
\cline{6-13}
& & & \multicolumn{2}{c|}{} & $512$ & 512 & $0$ & $55.52$ & $10.13$ & 0 & 50 & 100\\
\cline{6-13}
& & & \multicolumn{2}{c|}{} & $1024$ & 0 & $0$ & $8.21$ & $0$ & 0 & 25 & 13\\
\hline\hline

\multirow{2}{*}{L0-1} & \multirow{2}{*}{0.00} & \multirow{2}{*}{2.074} & \multicolumn{2}{|c|}{\multirow{2}{*}{0.822}} & $512$ & 0 & $0$ & $65.66$ & 0 & 0 &50 & 100\\
\cline{6-13}
& & & \multicolumn{2}{c|}{} & $1024$ & 0 & $0$ & $8.21$ & $0$ & 0 & 25 & 13\\
\hline\hline

\multirow{2}{*}{L0-2} & \multirow{2}{*}{0.00} & \multirow{2}{*}{2.053} & \multicolumn{2}{|c|}{\multirow{2}{*}{0.818}} & $512$ & 0 & $0$ & $65.66$ & 0 & 0 & 50 & 100\\
\cline{6-13}
& & & \multicolumn{2}{c|}{} & $1024$ & 0 & $0$ & $8.21$ & $0$ & 0 &  25 & 13\\
\hline\hline

\multirow{2}{*}{L0-3} & \multirow{2}{*}{0.00} & \multirow{2}{*}{2.00} & \multicolumn{2}{|c|}{\multirow{2}{*}{0.807}} & $512$ & $0$ & 0 & $65.66$ & 0 & 0 & 50 & 100\\
\cline{6-13}
& & & \multicolumn{2}{c|}{} & $1024$ & 0 & $0$ &  $8.21$ & $0$ & 0 & 25 & 13\\
\hline\hline

\multirow{2}{*}{L0-4} & \multirow{2}{*}{0.00} & \multirow{2}{*}{1.954} & \multicolumn{2}{c|}{\multirow{2}{*}{0.798}} & $512$ & $0$ & 0 & $65.66$ & 0 & 0 & 50 & 100\\
\cline{6-13}
& & & \multicolumn{2}{c|}{} & $1024$ & 0 & $0$ & $8.21$ & 0 & $0$ & 25 & 13\\
\hline\hline

L6 & 0.06 & 2.13 & 0.819 & 0.822 & $512$ & 0 & $512$ & $65.36$ & 0 & $29.57$ & 50 & 100\\
\hline\hline

\multirow{2}{*}{L10} & \multirow{2}{*}{0.10} & \multirow{2}{*}{2.13} & \multirow{2}{*}{0.809} & \multirow{2}{*}{0.815} & $512$ & 0 & $512$ & $65.16$ & 0 & $49.28$ & 50 & 100\\
\cline{6-13}
& & & & & $1024$ & 0 &$1024$ & $8.15$ & 0 &$6.16$ & 25 & 13\\
\hline\hline

\multirow{3}{*}{L15} & \multirow{3}{*}{0.15} & \multirow{3}{*}{2.13} & \multirow{3}{*}{0.798} & \multirow{3}{*}{0.806} & $512$ & 0 &$512$ & $64.92$ & 0 & $73.95$ & 50 &100\\
\cline{6-13}
& & & & & $512$ & 512 & $512$ & $54.78$ & $10.13$ & 73.95 & 50 & 100\\
\cline{6-13}
& & & & & $1024$ & 0 &$1024$ & $8.12$ & 0 &$9.24$ & 25 & 13\\
\hline
\end{tabular}}
\end{center}
\end{table*}

We use a subset of the \textsc{HADES} numerical simulations. We briefly describe the characteristics of this subset here and refer the reader to \citet{HADES} for further details. The simulations have been run using the TreePM+SPH code \textsc{Gadget-III} \citep{Springel_2005}. All simulations are run in a periodic box size of $1~h^{-1}{\rm Gpc}$ and all models share the value of the following cosmological parameters: $\Omega_{\rm m}=0.3175$, $\Omega_{\rm b}=0.049$, $\Omega_\Lambda=0.6825$, $n_s=0.9624$, $h=0.6711$, which are in very good agreement with those obtained by Planck \citep{Planck_2015}. A summary of the different simulations can be found in Table \ref{tab:i}.

\subsection{N-body}

In these simulations we follow the gravitational evolution of $N_{\rm cdm}$ CDM plus $N_\nu$ neutrino particles (only in the case of massive neutrinos models) from $z=99$ to $z=0$. For each cosmological model we have two types of simulations with different resolutions. In one case we have $N_{\rm cdm}=512^3$ and $N_\nu=512^3$ with 100 independent realizations for each model (low-resolution simulations) while in the other we have $N_{\rm cdm}=1024^3$ and $N_\nu=1024^3$ and 13 realizations for model (high-resolution simulations).  

The initial conditions are generated at $z=99$ using the \textit{rescaling} method outlined in \cite{Zennaro_2016} employing the Zeldovich approximation. The matter power spectra, transfer functions and growth rates required as input in \textit{reps}\footnote{https://github.com/matteozennaro/reps} are obtained through CAMB \citep{CAMB}. The gravitational softening is set to $1/40$ of the mean inter-particle distance for both CDM and neutrinos. The random seeds are the same for the same realization in different models and vary from realization to realization for the same model.

The simulations cover 8 different cosmological models, which can be split into models with massive and massless neutrinos. We simulate three different models with degenerate massive neutrinos with different masses: $M_\nu=0.06$ eV, $M_\nu=0.10$ eV and $M_\nu=0.15$ eV. Those simulations are run employing the so-called \textit{particle method} \citep{Brandbyge_2008, Viel_2010}, where neutrinos are described as a collisionless and pressureless fluid and therefore modeled as particles in the same fashion as CDM. 

\begin{figure}
\begin{center}
\includegraphics[width=0.49\textwidth]{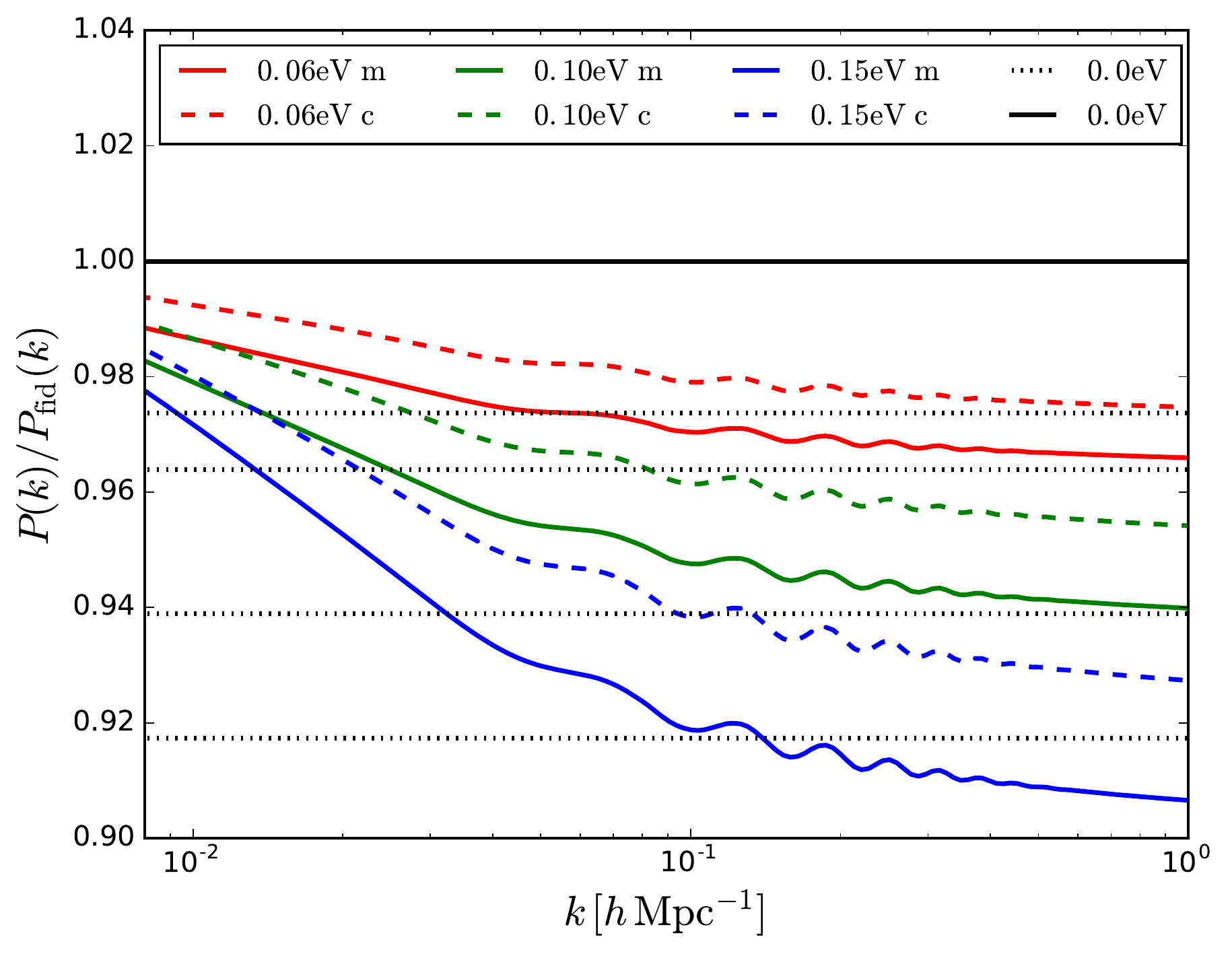}
\caption{We carry out our analysis using simulations for different cosmologies with massive and massless neutrinos. In the plot we show the linear power spectrum 
of each model, normalized by the power spectrum of the fiducial cosmology (solid black line), at $z=0$. The dotted black lines show the models with massless neutrinos and lower value of $\sigma_8$ than the fiducial model. The solid/dashed colored lines show the matter/CDM+baryons power spectrum of the models with massive neutrinos.}
\label{fig:Pk_lin}
\end{center}
\end{figure}

The different models with massless neutrinos only differ in the value of $\sigma_8$. We use those simulations to study the $M_\nu-\sigma_8$ degeneracy, i.e. to understand whether the difference between the models are driven by neutrino masses of by $\sigma_8$. In cosmologies with massive neutrinos $\sigma_8$ can be computed with respect to total matter, $\sigma_8^m$, or with respect to CDM+baryons, $\sigma_8^c$. It has been shown in several works (see e.g. \cite{Villaescusa-Navarro_2014, Castorina_2014, Ichiki-Takada, LoVerde_2014}) that in models with massive neutrinos, halos closely trace the CDM+baryon field, as opposed to the total matter (CDM+baryon+$\nu$) field.  The values of $\sigma_8$ in our simulations with massless neutrino simulations were chosen to match either $\sigma_8^c$ or $\sigma_8^m$ of the massive neutrino cosmologies. In Fig. \ref{fig:Pk_lin} we show the amplitude and shapes of the linear power spectra of the different models normalized by the linear power spectrum of the fiducial cosmology. 

\begin{figure*}
\begin{center}
\includegraphics[width=0.46\textwidth]{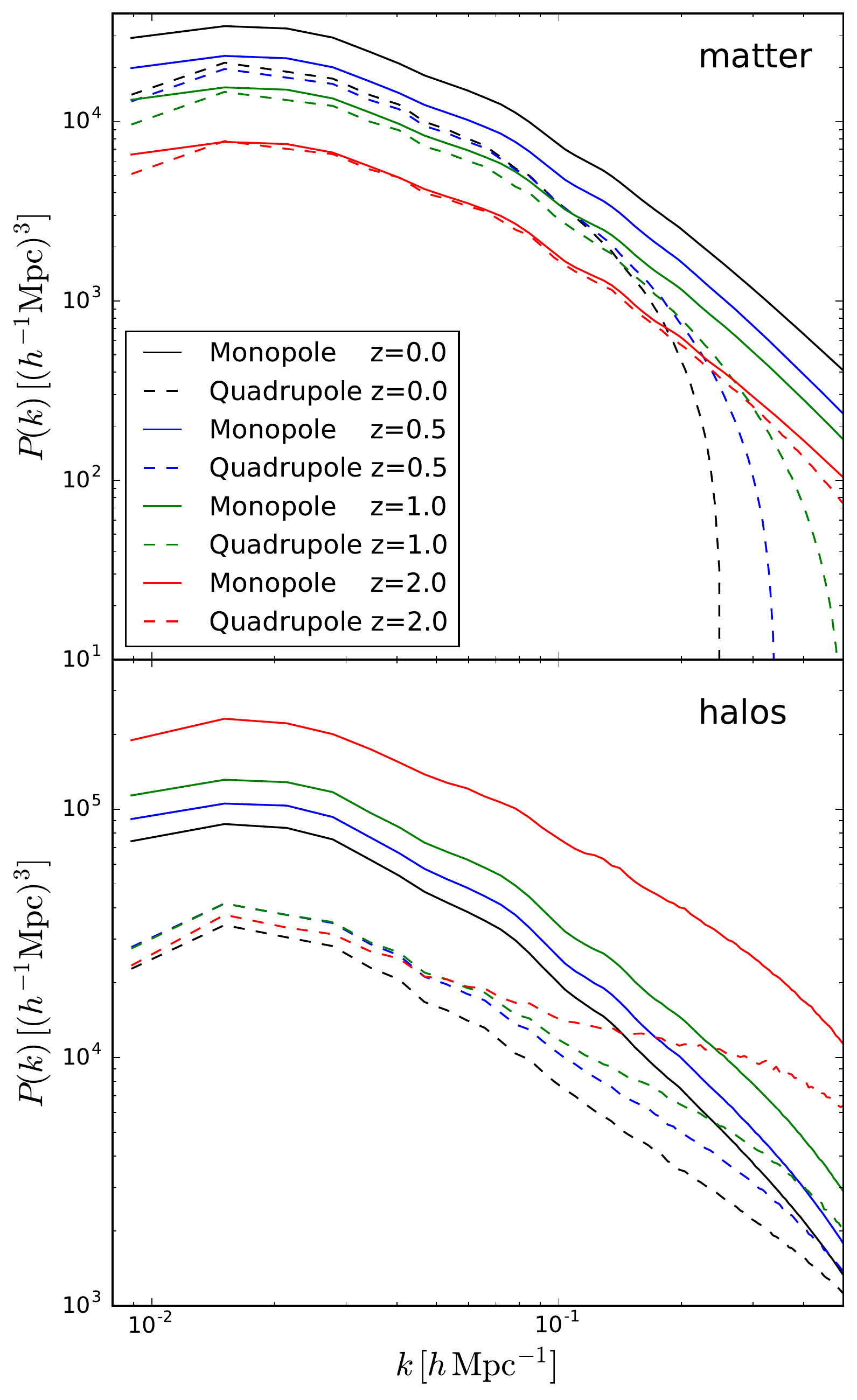}
\includegraphics[width=0.49\textwidth]{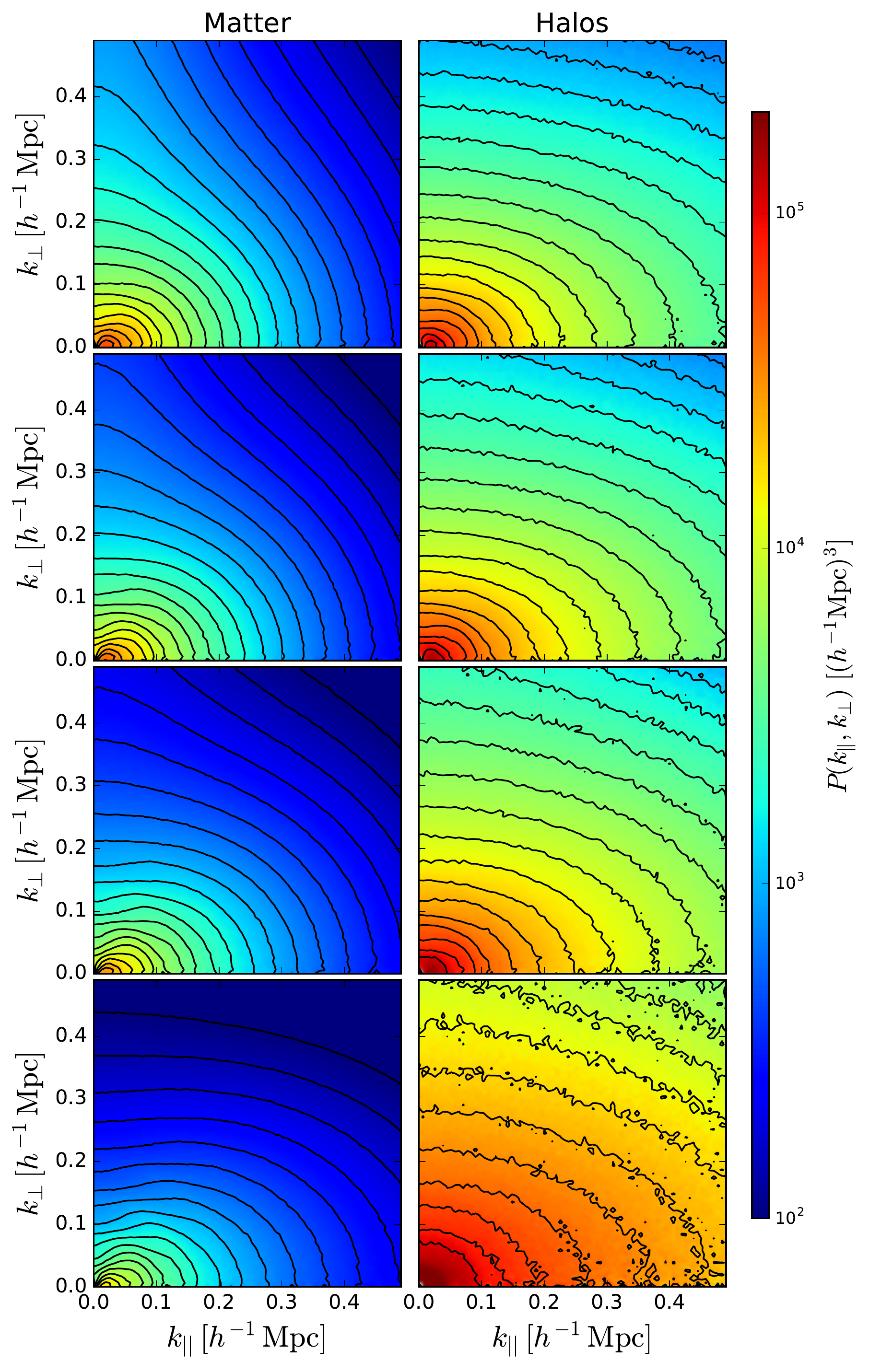}
\caption{The effects induced by the neutrino masses considered in this work are very small in all the quantities investigated. We thus focus our analysis on relative differences instead of absolute differences. As a reference, we show in this plot the absolute scale of the matter and halos power spectrum for the fiducial model with massless neutrinos. \textbf{Left:} Monopoles (solid lines) and quadrupoles (dashed lines) for matter (upper panel) and halos (bottom panel) at $z=0$ (black), $z=0.5$ (blue), $z=1$ (green) and $z=2$ (red). \textbf{Right:} 2D power spectra of matter (left column) and halos (right column) at $z=0$ (first row), $z=0.5$ (second row), $z=1$ (third row) and $z=2$ (fourth row). Black lines show isopower contours. We only consider halos with masses larger than $3.2\times10^{13}~h^{-1}M_\odot$.}
\label{fig:Pk_fid}
\end{center}
\end{figure*}

For each realization of each model we save snapshots at $z=0$, $z=0.5$, $z=1$, $z=2$ and $z=3$. Dark matter halos are identified by employing the Friends-of-Friends algorithm \citep{FoF} by setting the linking length parameter  $b=0.2$. The group-finder is run only on top of the CDM+baryon distribution, as the contribution of neutrino to the masses of halos has been shown to be negligible \citep{Villaescusa-Navarro_2011, Villaescusa-Navarro_2013,Brandbyge_2010,Ichiki-Takada,LoVerde_2014, Castorina_2014}. Only halos that contain more than 20 particles are identified, although for the main analysis we only consider halos with masses above $3.2\times10^{13}~h^{-1}M_\odot$ and $3\times10^{12}~h^{-1}M_\odot$ for the low- and high-resolution simulations, respectively. 

Since cosmic variance decreases with volume and is not affected by resolution, and given the fact that the effect induced by the neutrino masses considered in this paper are very small, we use for most of the analysis in this paper the lower-resolution $512^3$ simulations, and employ the higher-resolution ($1024^3$) simulations to compute the momentum bias for low-mass halos not resolved by the coarser simulations.

\subsection{Hydrodynamic}

Besides the N-body simulations, we also run hydrodynamic simulations for the fiducial and 0.15 eV models. In those simulations we follow the evolution of $512^3$ CDM + $512^3$ gas particles, plus $512^3$ neutrino particles for the 0.15eV model, in a box of $1~h^{-1}$Gpc from $z=99$ to $z=0$. The physical processes incorporated in those simulations are: radiative cooling by hydrogen and helium, uniform UV heating, star formation and supernova feedback following \citep{Springel-Hernquist_2003}. Supernova feedback is implemented as kinetic feedback with particles ejected at a velocity of $\simeq350$ km/s.

Initial conditions are also generated at $z=99$ using the Zeldovich approximation. A different power spectrum and growth rate is computed for each species, CDM, gas and neutrinos, using the \textit{reps} rescaling code (see \cite{Wessel_2016} for understanding the importance of generating the ICs properly for hydrodynamic simulations). For the gas particles we employ a softening length which is equal to the SPH gas radius. We do this to avoid the artificial coupling of CDM and gas at high redshifts \citep{Angulo_2013}. Thus, our simulations are able to reproduce the large-scale clustering of each individual component separately from linear theory (see appendix \ref{sec:sph} for a discussion on the impact of the gas smoothing lengths on the clustering of matter and halos in real- and redshift-space). 

We save snapshots at redshifts 0, 0.5, 1, 2 and 3 and identify halos through FoF. Halos are identified using both CDM and baryons. For the analysis we only take halos with masses larger than $3.2\times10^{13}~h^{-1}M_\odot$.

We use these simulations to investigate the impact of baryons and astrophysical effects on our results.

\section{General features}
\label{sec:features}

In this work we are interested in investigating the impact of neutrino masses on several quantities, such as the clustering of matter, CDM+baryons and halos in redshift-space or the growth rate. We will focus our analysis on scales $k\simeq[10^{-2},0.5]~h{\rm Mpc}^{-1}$ and redshifts $z\in[0,2]$. While the amplitudes of the monopoles, quadrupoles and the fully 2D power spectra varies several orders of magnitude in that redshift range, the relative differences among the models we study here are limited to a few percent. What this means in practice is that if we plot, let's say, the monopoles at a given redshift for all the different models we will see one line on top of each other. Similar arguments apply to the case of the growth rate. 

For this reason in this paper we focus our analysis to relative differences instead of absolute differences. In this section we show the monopoles, quadrupoles and fully 2D power spectrum of the fiducial model, together with the growth rate of the model with 0.15 eV neutrinos. 

\subsection{Clustering}

For each of the 100 realizations of the fiducial cosmology we have computed the monopoles, quadrupoles and fully 2D power spectra of both the matter and halos at redshifts 0, 0.5, 1 and 2 and what we show here the mean from all realizations. We have subtracted the $\bar n^{-1}$ shot-noise of the halo field when computing the monopoles and 2D power spectra for each realization and redshift.

In the left panels of Fig. \ref{fig:Pk_fid} we show the monopoles and quadrupoles of the matter and halos distributions in redshift-space at redshifts 0, 0.5, 1 and 2. For the matter distribution the amplitude of the monopoles increases with redshift, due to the non-linear growth of the matter perturbations. For halos the trend is inverted due to the fact that we are consider halos with the same minimum mass in all cases, $3.2\times10^{13}~h^{-1}M_\odot$: the halo bias increases with redshift more rapidly that perturbations growth  (see appendix \ref{sec:halo_bias}). 

\begin{figure}
\begin{center}
\includegraphics[width=0.49\textwidth]{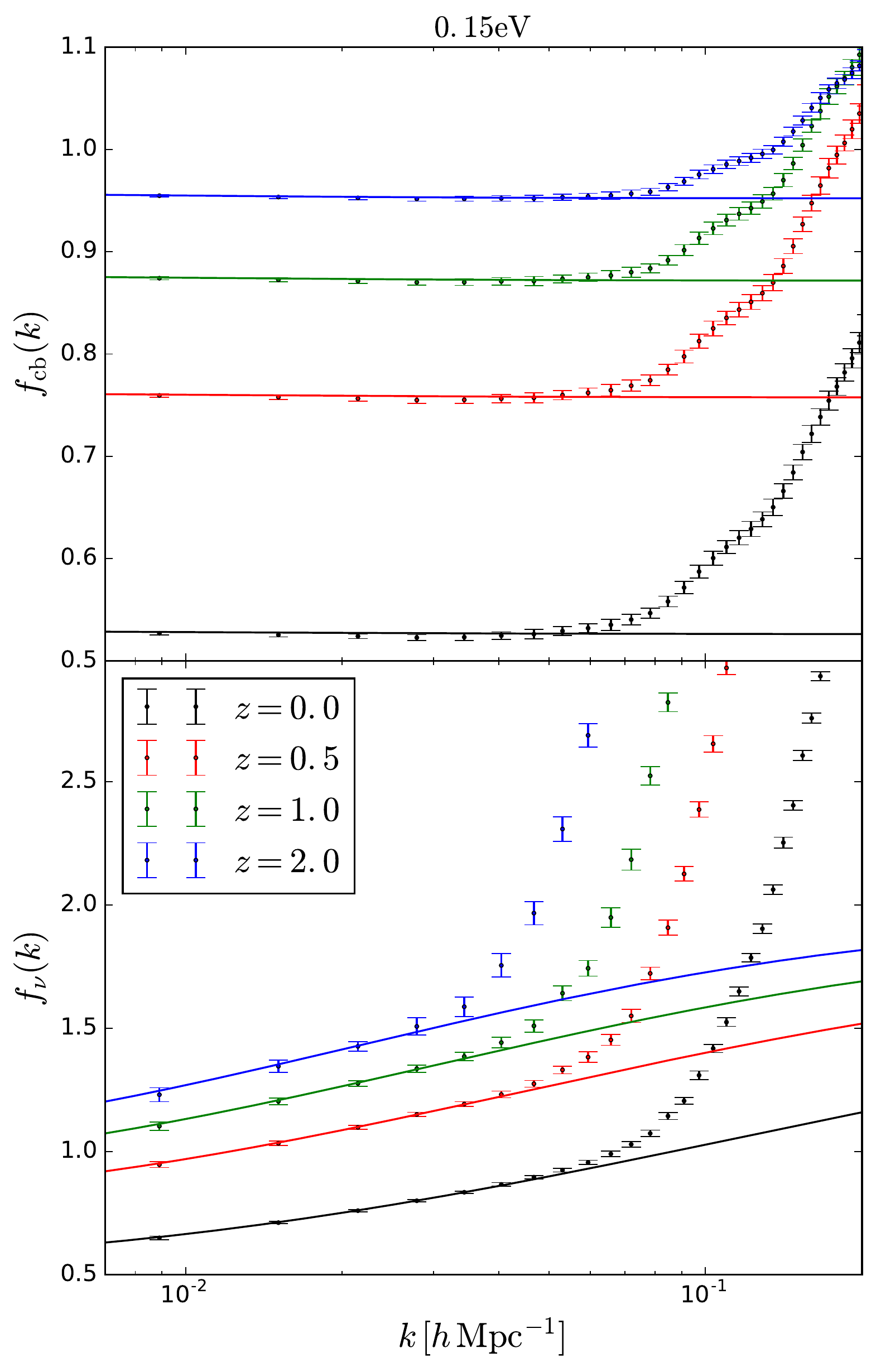}
\caption{This plot shows the absolute growth rate of the CDM+baryons (top) and neutrino (bottom) fields at $z=0$ (black), $z=0.5$ (red), $z=1$ (green) and $z=2$ (blue) for the model with 0.15 eV neutrinos. The points with error bars show the results from our simulations while the solid lines represent the predictions from linear theory. Deviations from linear theory are due to non-linearities, the estimator and the presence of shot-noise (only in the case of neutrinos).}
\label{fig:growth_absolute}
\end{center}
\end{figure}

We find a cutoff in the quadrupole of the matter distribution on small scales induced by the Fingers-of-God (FoG), an effect whose amplitude increases with decreasing redshift. At high redshifts and on large scales the amplitude of the quadrupole and monopoles of the matter distribution agrees very well. This is a consequence of the linear growth rate being close to 1 (see Fig. \ref{fig:growth_absolute}), i.e. $1+2f/3+f^2/5\sim4f/3+4f^2/7$, where $f$ is the growth rate.

The halos quadrupole do not present instead a cutoff on small scales, reflecting the almost absence of FoG in the halo field. The amplitude of the halos quadrupole on large-scales shows a non-monotonic relation with redshift, that can be explained by taking into account the different time evolution of the growth factor, growth rate and halo bias. In the linear regime, the amplitude of the quadrupole goes as $D^2b^2(4\beta/3+4\beta^2/7)$, where $D$ is the growth factor, $b$ is the halo bias and $\beta=f/b$. While both $b$ and $f$ increase with redshift, $D$ shrink with it. 

In the right part of Fig. \ref{fig:Pk_fid} we show the 2D power spectrum for matter and halos fields in redshift-space at redshifts 0, 0.5, 1 and 2. The power spectrum of the matter field shows the increase of power with decreasing redshift displayed in the left part of Fig. \ref{fig:Pk_fid}. From the structure of the 2D matter power spectrum it is also clear that the magnitude of the FoG decreases with redshift. 

The 2D power spectrum of the halo field shows the increase of power with redshift that we found in the halos monopole, due to the larger halo bias of the halos. The lack of FoG in the halo field is clearly reflected in the structure of the halos 2D power spectrum. The number density of halos of fixed mass decreases with redshift. This is the reason why the 2D halo power spectrum becomes noisier with redshift. 

\subsection{Growth rate}

$\vec{j}$

We estimate the growth rate of both the CDM+baryons and neutrino fluids as
\be
f_{\rm i}(k,a)=\frac{1}{aH}\sqrt{\frac{P_{\rm P_{\rm i}(k)}}{P_{\delta_{\rm i}(k)}}}~,
\label{eq:growth}
\ee
where $i\in[{\rm cb},\nu]$ and $P_{\rm i}=(\vec{\nabla}\cdot \vec{J})_{\rm i}=\vec{\nabla}\cdot[(1+\delta)\vec{v}]_{\rm i}$ is the divergence of the momentum of the field ${\rm i}$. $P_{\rm P_{\rm i}}(k)$ and $P_{\delta_{\rm i}}(k)$ are the divergence of the momentum and density power spectrum, respectively. On linear scales, the above growth factor reduces to the standard linear growth rate, while on small scales it measures an angle-average growth rate as we show below.

We can express the overdensity of the species ${\rm i}$ at any time as $\delta_{\rm i}(\vec{k},a)=D_{\rm i}(\vec{k},a)/D_{\rm i}(\vec{k},a_0)\delta_{\rm i}(\vec{k},a_0)$, and using the continuity equation we get
\be
aHf_{\rm i}(\vec{k},a)\delta_{\rm i}(\vec{k},a)=P_{\rm i}(\vec{k},a)~,
\ee
where $f_{\rm i}(k,a)=d\log D_{\rm i}(k,a)/d\log a$. The relation between the growth rate we estimate from Eq. \ref{eq:growth} and $f_{\rm i}(\vec{k},a)$ is then given by
\be
P_{\delta_{\rm i}}(k,a)f^2_{\rm i}(k,a)=P_{\delta_{\rm i}f_{\rm i}}(k,a)
\ee
with the right hand side representing the power spectrum of the product of $\delta_{\rm i}$ and $f_{\rm i}$.

We have computed the growth rate of the CDM+baryon and neutrino fields for each realization (100 in total) of the model with 0.15 eV neutrinos and in Fig. \ref{fig:growth_absolute} we show the average and standard deviation at different redshifts. We show the results for this model and not for the fiducial model because in terms of the growth rate of the CDM+baryon field results are very similar but we can also show the results for the neutrino field. 

\begin{figure}
\begin{center}
\includegraphics[width=0.49\textwidth]{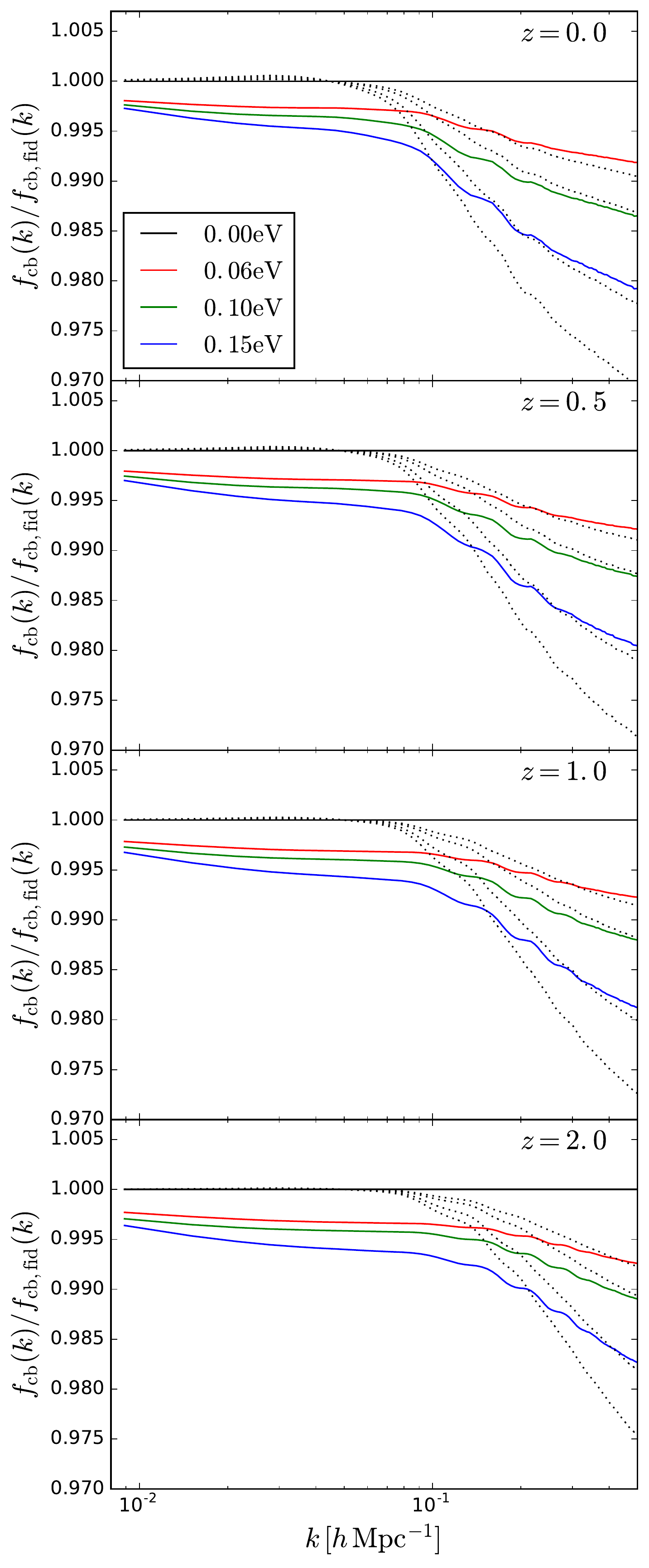}
\caption{Growth rate of the CDM+baryon field, normalized by the growth rate of the fiducial model, at $z=0$ (first), $z=0.5$ (second), $z=1$ (third) and $z=2$ (fourth). Red, green and blue lines show the results for the models with 0.06 eV, 0.10 eV and 0.15 eV, respectively, while the dotted lines display the results for models with massless neutrinos and different values of $\sigma_8$ (see Fig. \ref{fig:Pk_lin}). The amplitude of the growth rate on large scales is sensitive to neutrino masses while it does not depend on $\sigma_8$. $0.15$ eV neutrinos change the amplitude of the growth rate by less than $\simeq2\%$ down to $k=0.5~h^{-1}$Mpc. On very small scales, neutrino masses and $\sigma_8$ produce similar effects.}
\label{fig:growth_CDM}
\end{center}
\end{figure}

We find that linear theory is able to reproduce very well the results of the simulations on large scales for both CDM+baryons and neutrinos at all redshifts.
We notice that even at linear order the growth of both CDM+baryons and neutrinos is scale-dependent, in contrast with the growth of matter for a massless neutrino cosmology\footnote{Notice that this is only true on scales below the horizon.}. From Fig. \ref{fig:growth_absolute} this effect is clear for neutrinos, but it is so small for CDM+baryons that it can not be  properly seen in the plot. This effect can however be visualized clearly in Fig. \ref{fig:growth_CDM}.

On smaller scales the growth rate measured in simulations depart, as expected, from linear theory. There are several reasons for this: 1) the growth rate becomes non-linear, 2) the quantity that we measure in simulations on small scales is not exactly the growth rate, but a convolution of density perturbations and growth rate, 3) in the case of neutrinos there is a non-negligible contribution coming from the neutrinos shot-noise that increases with redshift. The last point explains why the agreement with linear theory improves on smaller scales for higher redshift for CDM+baryons, but get worse for neutrinos. Results for other models, and the comparison with linear theory, is as good for the others models as for the model with 0.15eV.

\section{Matter distribution}
\label{sec:matter}

\begin{figure*}
\begin{center}
\includegraphics[width=0.93\textwidth]{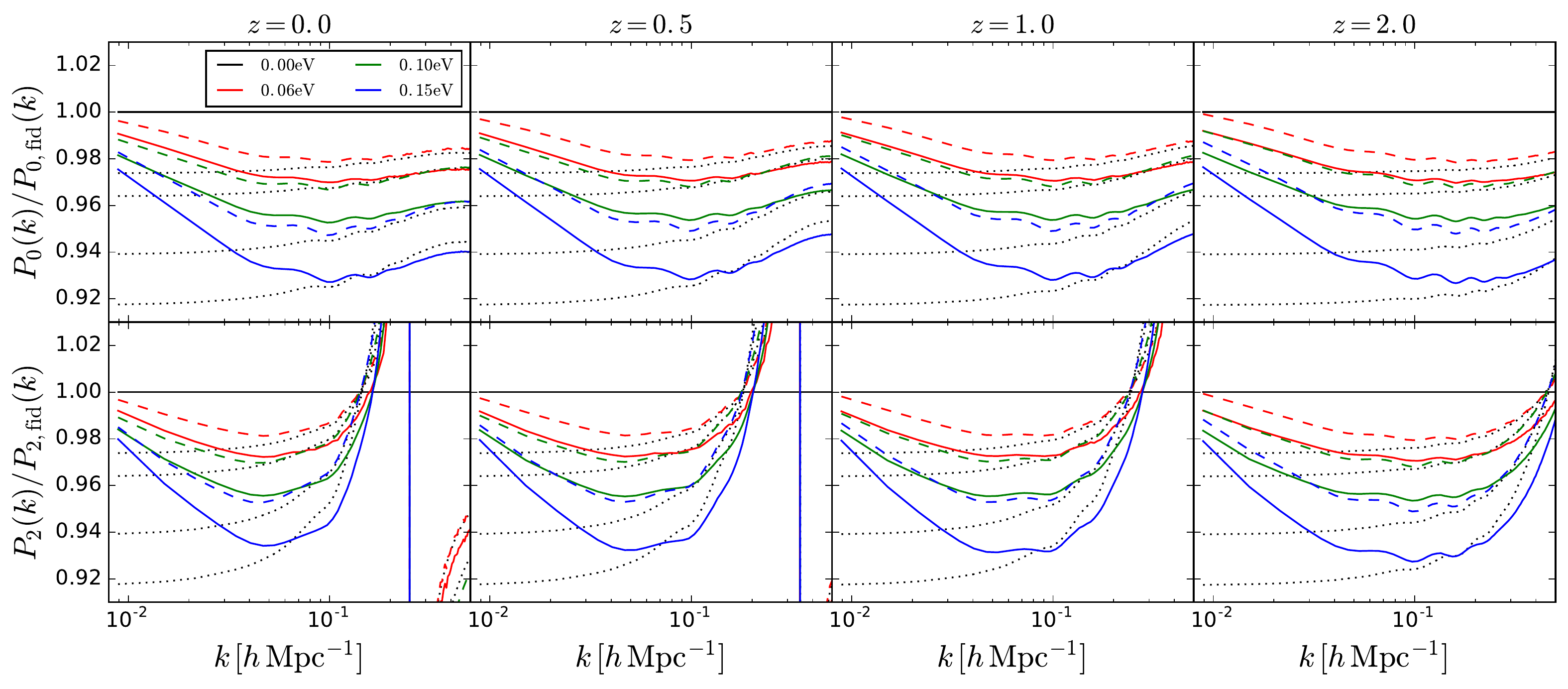}\\
\vspace{0.8cm}
\includegraphics[width=0.93\textwidth]{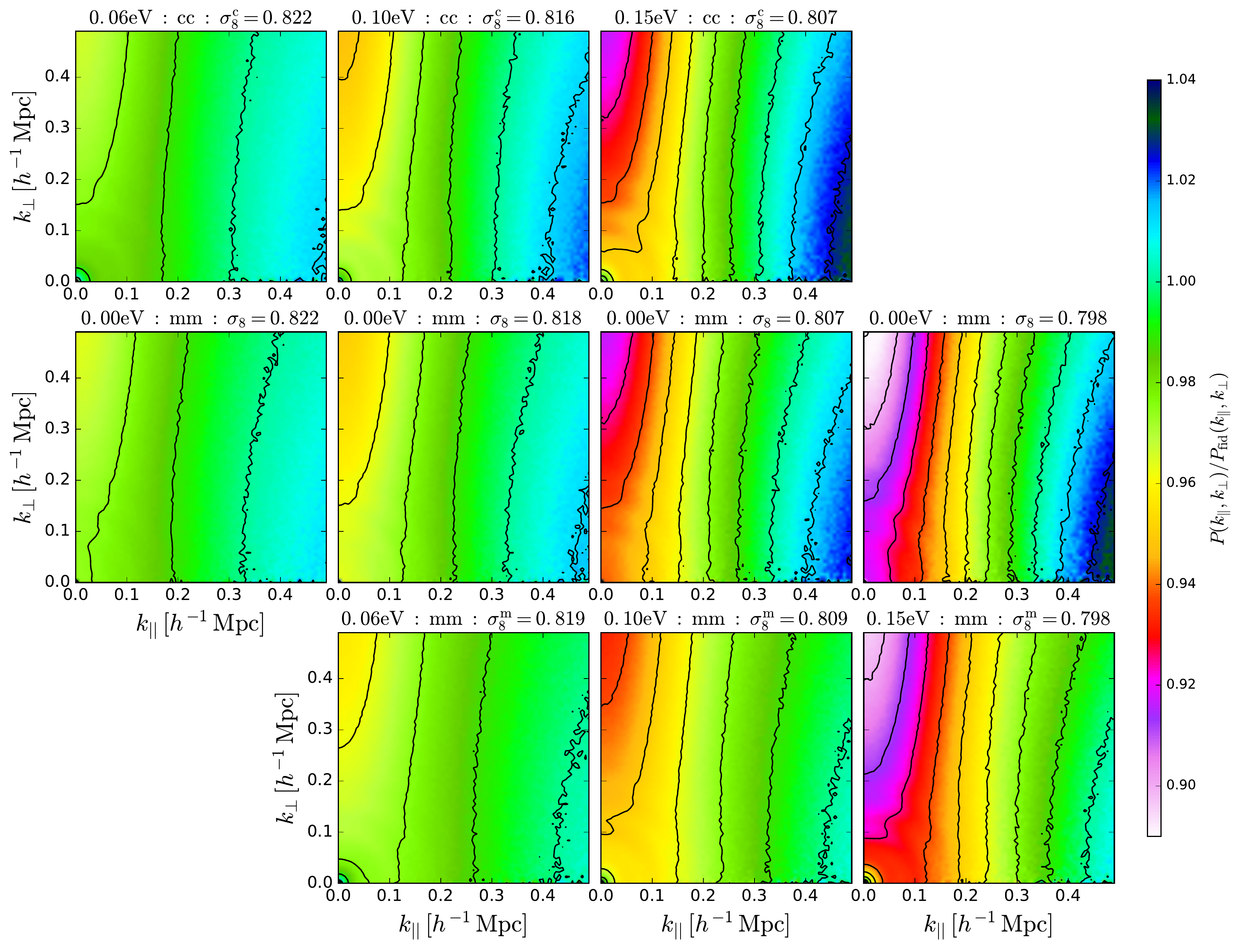}
\caption{Impact of neutrino masses and $\sigma_8$ on the clustering of matter in redshift-space. \textbf{Top:} Monopoles (upper row) and quadrupoles (bottom row) for the different models, normalized by the results of the fiducial cosmology, at $z=0$ (left), $z=0.5$ (middle-left), $z=1$ (middle-right) and $z=2$ (right). The dotted black lines show the results for the models with massless neutrinos and lower values of $\sigma_8$ than the fiducial model (see Fig. \ref{fig:Pk_lin}). The red, green and blue lines display the results for the models with 0.06 eV, 0.10 eV and 0.15 eV neutrinos, respectively. Solid and dashed lines represent the results for the matter and CDM+baryon fields, correspondingly. \textbf{Bottom:} 2D power spectrum of the different models, normalized by the results of the fiducial model at $z=0$. The top part of each panel shows the model for which the results are displayed. For models with massive neutrinos we show both the 2D power spectrum of CDM+baryons (upper row) and total matter (bottom row). Panels in the same column have similar values of $\sigma_8$. The solid black lines show isopower contours. While the effect of neutrinos is very different to the one of $\sigma_8$ on large scales, on small scales there is an almost perfect degeneracy between these parameters for the CDM+baryons field. In the matter field that degeneracy is less prominent and the quadrupole can be used to break it.}
\label{fig:matter}
\end{center}
\end{figure*}

In this section we investigate the impact of neutrino masses on redshift-space distortions in the matter and CDM+baryon fields. For each realization and redshift of each model we have computed the matter and CDM+baryon 3D power spectra in real-space\footnote{We do not show these results here as they have been extensively studied in other works.}, the monopoles, quadrupoles and the 2D power spectra in redshift-space. Here we show the results for the mean, i.e. for a volume equal to $\simeq100~(h^{-1}{\rm Gpc})^3$.

In the upper panels of Fig. \ref{fig:matter} we show the monopoles and quadrupoles in redshift-space, normalized by the results of the fiducial cosmology, for matter and CDM+baryons, for the different models,. The effect of massive neutrinos on the matter or CDM+baryons monopole in redshift-space exhibits the same features as in real-space: a suppression on power from large to small scales that increases with $k$ and an increase in power on very small scales. This produces the characteristic \textit{spoon-shape} that neutrinos induce in real-space \citep{Brandbyge_2008,Viel_2010,Massara_2014}. The effect of varying $\sigma_8$ on the monopole for models with massless neutrinos is a suppression on power that increases on large scales. As can be seen, the effect of $\sigma_8$ and $M_\nu$ is not degenerate on scales larger than $\simeq0.1~h{\rm Mpc}^{-1}$, while on smaller scales the effect of both parameters are very degenerate. We also notice that the different time evolution of the models with massive and massless neutrinos partially break that degeneracy. 

The quadrupoles exhibit a similar behavior to the monopoles. A more pronounced spoon-shape is found in the ratio between models with massive neutrinos to the fiducial case. On large scales, the quadrupoles induced by $M_\nu$ differ in shape to those produced by $\sigma_8$, while on small scales we find that the effect of neutrinos is very degenerate with $\sigma_8$ only for the CDM+baryon field. Interestingly, our results point out that at the level of total matter, the quadrupoles of models with massless neutrinos can be distinguished from those with massive neutrinos on both large and small scales.

In the bottom part of Fig. \ref{fig:matter} we show the fully 2D power spectrum for the different models, normalized to the 2D power spectrum of the fiducial model, in redshift-space at $z=0$. As can be seen both $\sigma_8$ and $M_\nu$ changes the structure of the 2D power spectrum in a complicated way. As we have seen from the monopoles and quadrupoles results, the effect of $M_\nu$ on large-scales  is different to the one of $\sigma_8$: the reduction of power on large-scales from massive neutrinos is smaller than the one induced by decreasing $\sigma_8$, as can be seen from the bottom-left part of the panels showing results for models with massive neutrinos. In agreement with our previous results for monopoles and quadrupoles, we find a strong degeneracy between $M_\nu$ and $\sigma_8$ on small scales, that it is more prominent for CDM+baryons than for total matter. 

To verify the magnitude of the $M_\nu$-$\sigma_8$ degeneracy in the CDM+baryon field and its almost independence on $\mu$ for large wavenumbers, we have also computed the variance of the field
\be
\sigma^2_P(k)=\int_0^1 P^2(k,\mu)d\mu-\left(\int_0^1P(k,\mu)d\mu\right)^2
\ee
for the different models. We find a very strong degeneracy also when employing this quantity, confirming quantitatively our visual impression from Fig. \ref{fig:matter}.

\begin{figure*}
\begin{center}
\includegraphics[width=1.0\textwidth]{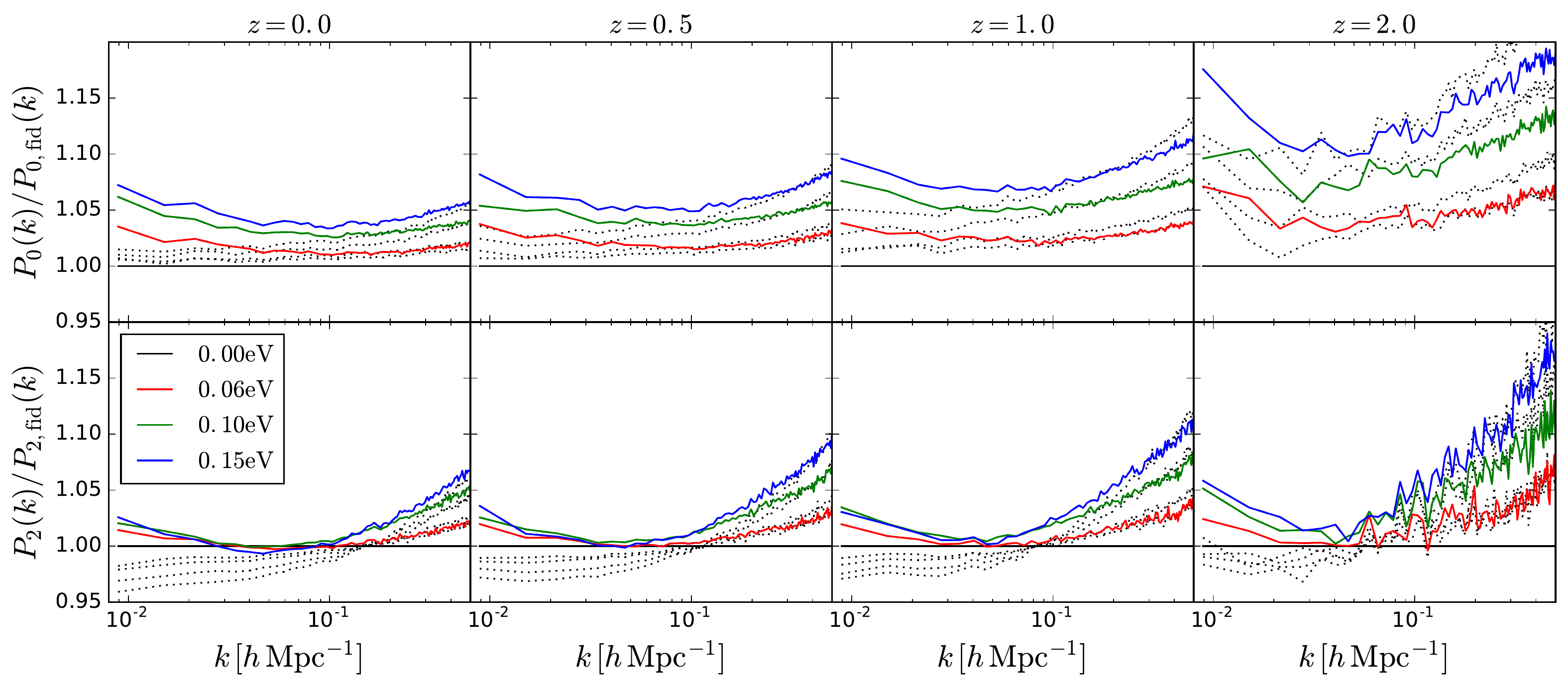}\\
\vspace{0.8cm}
\includegraphics[width=1.0\textwidth]{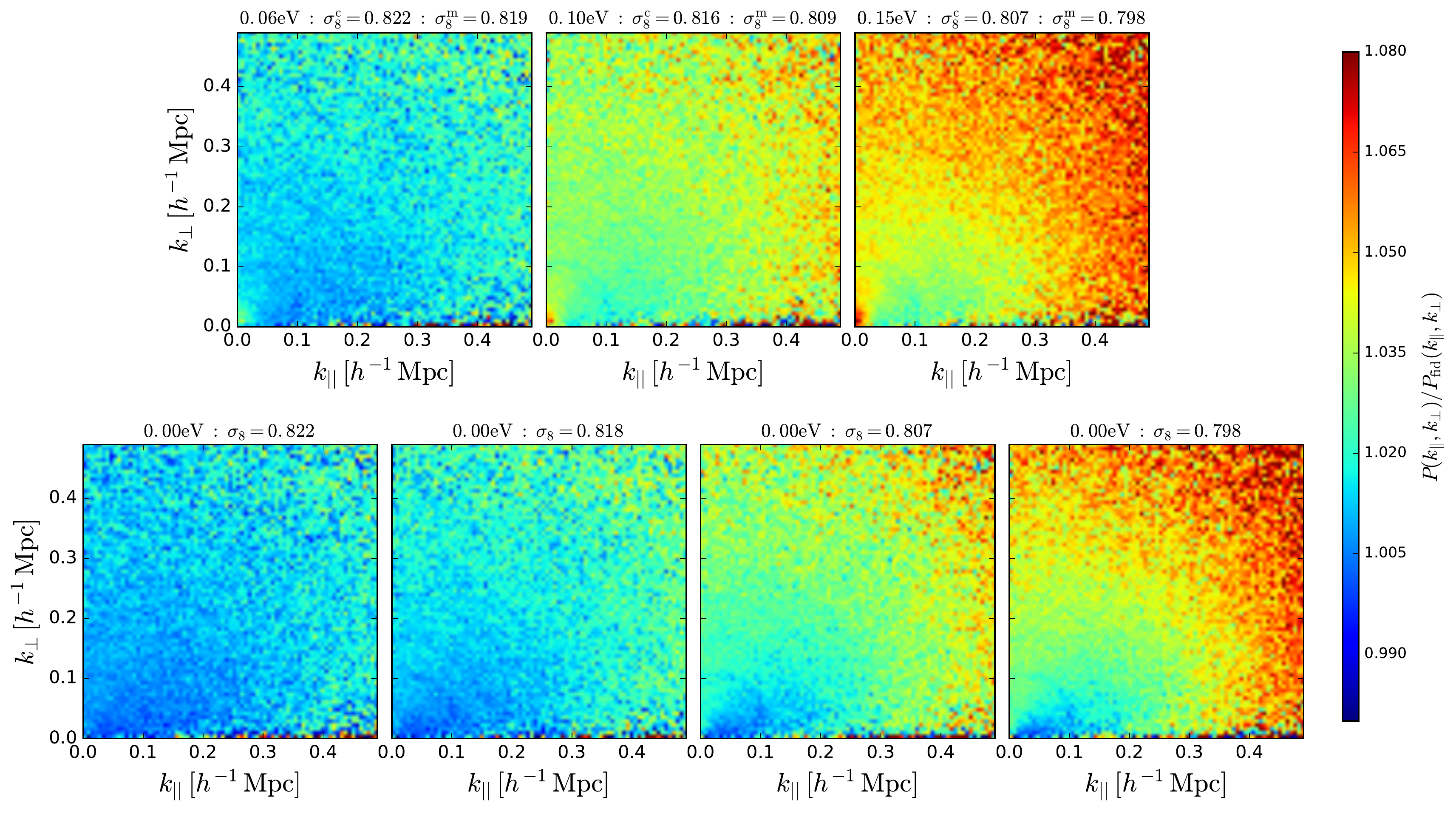}
\caption{Impact of neutrino masses and $\sigma_8$ on the clustering of halos with masses above $3.2\times10^{13}~h^{-1}$Mpc in redshift-space. \textbf{Top:} Monopoles (upper row) and quadrupoles (bottom row) for the different models, normalized by the results of the fiducial cosmology, at $z=0$ (left), $z=0.5$ (middle-left), $z=1$ (middle-right) and $z=2$ (right). The dotted black lines show the results for the models with massless neutrinos and lower value of $\sigma_8$ than the fiducial model (see Fig. \ref{fig:Pk_lin}). The red, green and blue lines display the results for the models with 0.06 eV, 0.10 eV and 0.15 eV neutrinos, respectively. \textbf{Bottom:} Ratio between the 2D power spectrum of halos for the different models to the results of the fiducial model. The effects of $M_\nu$ and $\sigma_8$ on the halo clustering on large scales is very different, while on small scales scales they are more degenerate, but much less than for the CDM+baryons of matter fields.}
\label{fig:halos}
\end{center}
\end{figure*}

\subsection{Growth rate}
\label{subsec:growth}

As we shall see below, an important quantity to describe the clustering properties of halos or galaxies is the growth rate of the CDM+baryon fluid. We, for the first time, measure this quantity in cosmologies with massive neutrinos and investigate the impact of neutrino masses and $\sigma_8$ on its amplitude and shape in the fully non-linear regime.

In Fig. \ref{fig:growth_CDM} we show the growth rate for the different cosmological models, normalized by the growth rate of the fiducial model, at redshifts 0, 0.5, 1 and 2. Results represent the mean over the 100 realizations of each model. We find that the growth rate of the CDM+baryon fluid is, on all scales and redshifts probed by our simulations, smaller than the fiducial model with massless neutrinos. The growth becomes slower as the sum of the neutrino masses increases, in agreement with the expectations of linear theory. For the neutrino masses considered here, the difference in the growth rate with respect to the fiducial model is never larger than $\sim2\%$ down to $k=0.5~h{\rm Mpc}^{-1}$. 

The dotted lines in Fig. \ref{fig:growth_CDM} show the results for the models with massless neutrinos and different values of $\sigma_8$. As expected from linear theory, on large scales the growth rate does not exhibit any dependence with $\sigma_8$, while on small scales the non-linear gravitational evolution induces a scale-dependent different with respect to the fiducial model. We find a degeneracy, on small scales, between the growth rate of models with massless neutrinos sharing the same value of $\sigma_8$ as the models with massive neutrinos.

\section{Halos distribution}
\label{sec:halos}

We now investigate the impact of massive neutrinos and $\sigma_8$ on the clustering properties of dark matter halos in redshift-space. We have computed, for each realization of each model, the halos monopoles, quadrupoles and the 2D power spectrum in redshift-space, subtracting the contribution from shot-noise. The results we show here represent the average over all realizations for each model. We only consider the simulations with low resolution and therefore focus our analysis on halos with masses above $3.2\times10^{13}~h^{-1}M_\odot$. 

In the upper part of Fig. \ref{fig:halos} we show the ratios between the monopoles and quadrupoles of the different models, to the monopole and quadrupole of the fiducial model, at redshifts 0, 0.5, 1 and 2 in redshift-space. In contrast to the results of matter and CDM+baryons, the amplitude of the halos monopole is larger, on all the scales considered here, than that of the fiducial model. The reason for this is that we are considering halos of fixed mass, so the lower value of $\sigma_8$ in the models with massive neutrinos translate into a larger halo bias (see appendix \ref{sec:halo_bias}) that boost the overall amplitude of the monopole. The same reason also explains while the amplitude of the monopoles increase with decreasing $\sigma_8$ for models with massless neutrinos. On large scales, the effect of massive neutrinos on the halo monopole is not degenerate with the effect induced by $\sigma_8$, similarly to what happens in the matter and CDM+baryon fields. We find however a degeneracy between $M_\nu$ and $\sigma_8$ on small scales, although it is less strong than the one present for matter and CDM+baryons.

We find that the effect of massive neutrinos of the halo quadrupole on large scales is also pretty different to the one induced by $\sigma_8$, in agreement with our results for the matter and CDM+baryon fields. Our results point out that a degeneracy between $\sigma_8$ and $M_\nu$ is present on small scales, that is weaker for models with larger neutrino masses. We however emphasize that even with 100 simulations per model the error bars on the halo quadrupoles are too large to reach robust conclusions. 

In the bottom part of Fig. \ref{fig:halos} we display the ratio between the 2D power spectrum of halos to the results of the fiducial model in redshift-space, for the different models, at $z=0$. As in the case of matter and CDM+baryons, we find that both massive neutrinos and $\sigma_8$ induce a non-trivial effect on the structure of the 2D power spectrum. On large scales, the modifications to the structure of the 2D power spectrum induced by $M_\nu$ and $\sigma_8$ are different, as expected from our previous results. On small scales both effect are however very degenerate. We find that the degeneracy is much stronger when the value of $\sigma_8$ of the model with massless neutrinos matches the value of $\sigma_8^c$ of the cosmology with massive neutrinos, in agreement with previous works \citep{Castorina_2014, Villaescusa-Navarro_2014, Castorina_2015}.

It is interesting to notice that massive neutrinos induce a feature on the 2D power spectrum on large scales that it is not reproduced by the models with massless neutrinos and different $\sigma_8$: the 2D power spectrum exhibits a larger variance as a function of $\mu$ for models with massive neutrinos than for massless models. This can be seen in the bottom part of Fig. \ref{fig:halos} as an excess of power in the perpendicular direction with respect to the line of sight. This effect is however easily explained by linear theory and it is just a consequence of the different halo bias, linear matter power spectrum and growth rate between the models with massless and massive neutrinos.

\subsection{Momentum bias}

In this subsection we try to answer the question whether, in cosmologies with massive neutrinos, the peculiar velocities of dark matter halos follow that of the underlying CDM+baryon fluid. In other words, whether there is a velocity bias between the halos and CDM+baryons. The root of this question resides in the fact that, in models with massive neutrinos, there are two different fields with different clustering properties: the total matter field and the CDM+baryon field. In \cite{Villaescusa-Navarro_2014, Castorina_2014} it was shown for the first time that the clustering properties of halos are dictated by the properties of the underlying CDM+baryon field rather than by the total matter field. It is also important as the theory template used to extract information on neutrino masses from velocity/momentum observations, as kSZ surveys, depend on it.

Employing similar arguments as those described in \cite{Castorina_2014}, it seems natural to expect that the statistical properties of the halo velocity field on large scales should resemble that of the underlying CDM+baryon field. The purpose of this section is to investigate whether the velocity bias between halos and CDM+baryons differs from 1 on large scales. 

The velocity bias can de defined as
\be
b^2_{\rm v}(k)=\frac{P_{\theta_{\rm h}}(k)}{P_{\theta_{\rm c}}(k)}
\ee
where $\theta=-\vec{\nabla}\cdot\vec{v}$. We have tried to measure this quantity directly from our simulations by estimating the velocity field either by assigning velocities from particles/halos to a grid using the cloud-in-cell (CIC) or the piecewise-cubic-spline (PCS) interpolation schemes \cite[see e.g.][]{Sefusatti_2016} or using the Delaunay tessellation field interpolation (DTFE) method \citep{Cautun_2011}. While for the CDM+baryon field the usage of the two different methods yield similar results, for halos the situation is very different. We find that in order to avoid having empty cells where the velocity is not well defined, a very coarse grid need to be employed with either CIC or PCS \citep{Pueblas_2009}. That coarseness avoids a reliable estimation of the halo velocity power spectrum even on large scales, due to effects such as aliasing. On the other hand, by employing the DTFE method, where the halo velocity field is well defined everywhere, we find a very strong sampling bias that depends on the number density of the tracers, in agreement with previous works \cite[see e.g.][]{Jennings_2014, Zheng_2015}. 

\begin{figure*}
\begin{center}
\includegraphics[width=0.87\textwidth]{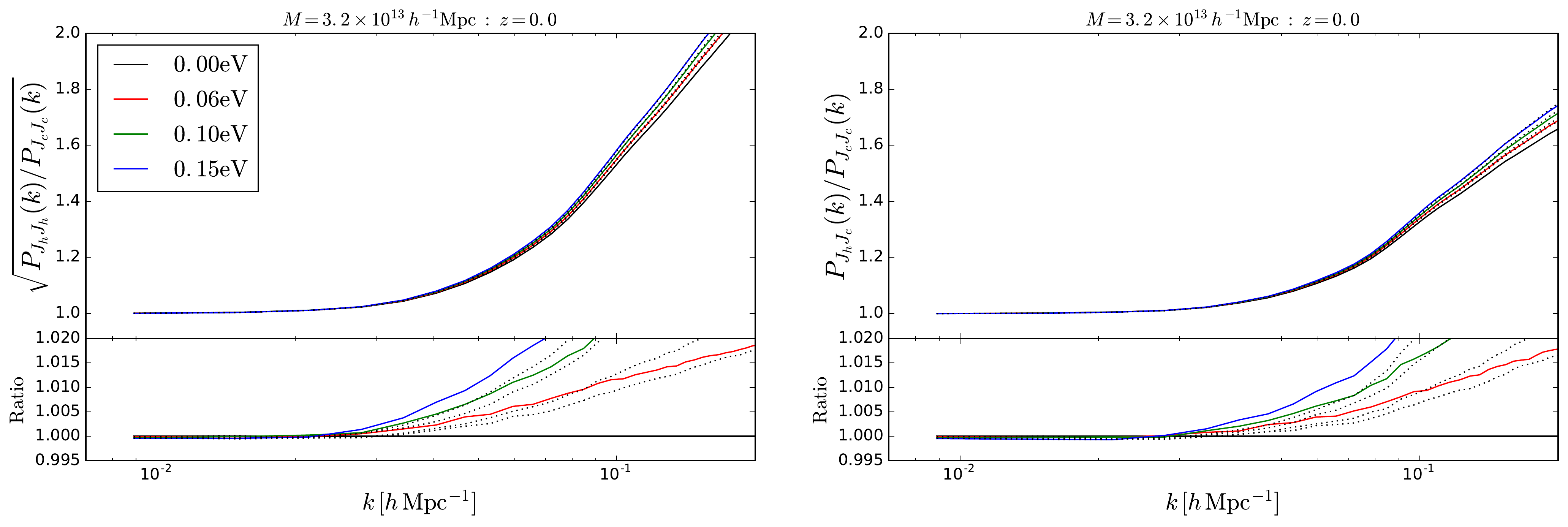}
\includegraphics[width=0.87\textwidth]{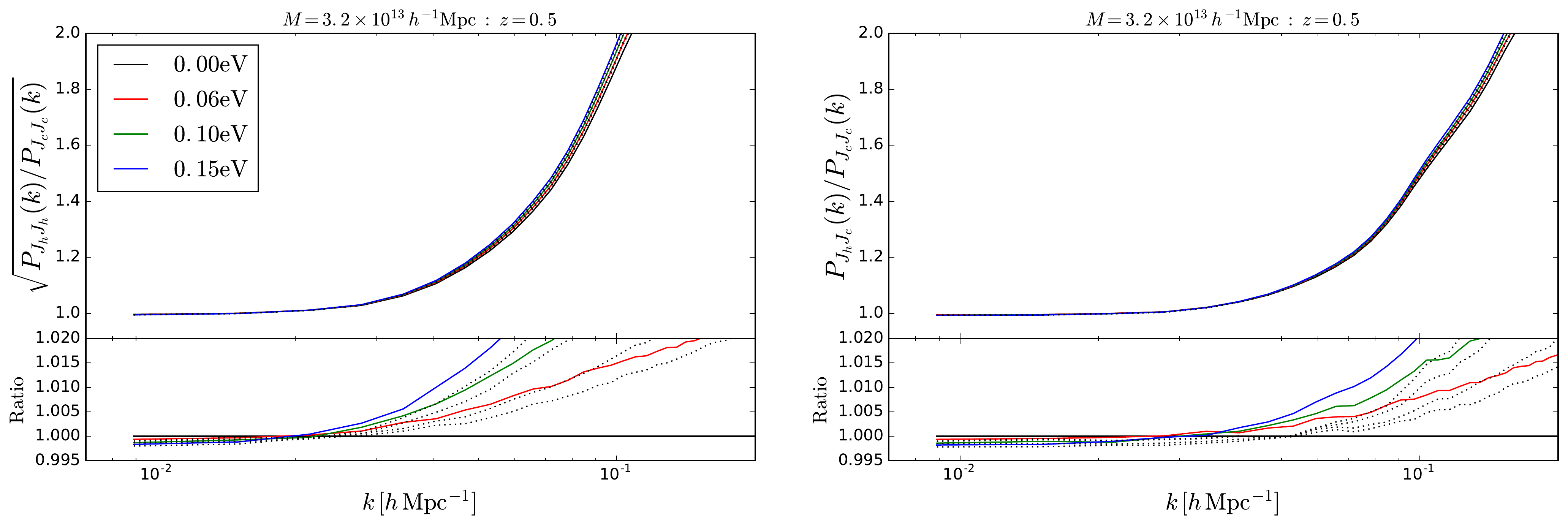}
\includegraphics[width=0.87\textwidth]{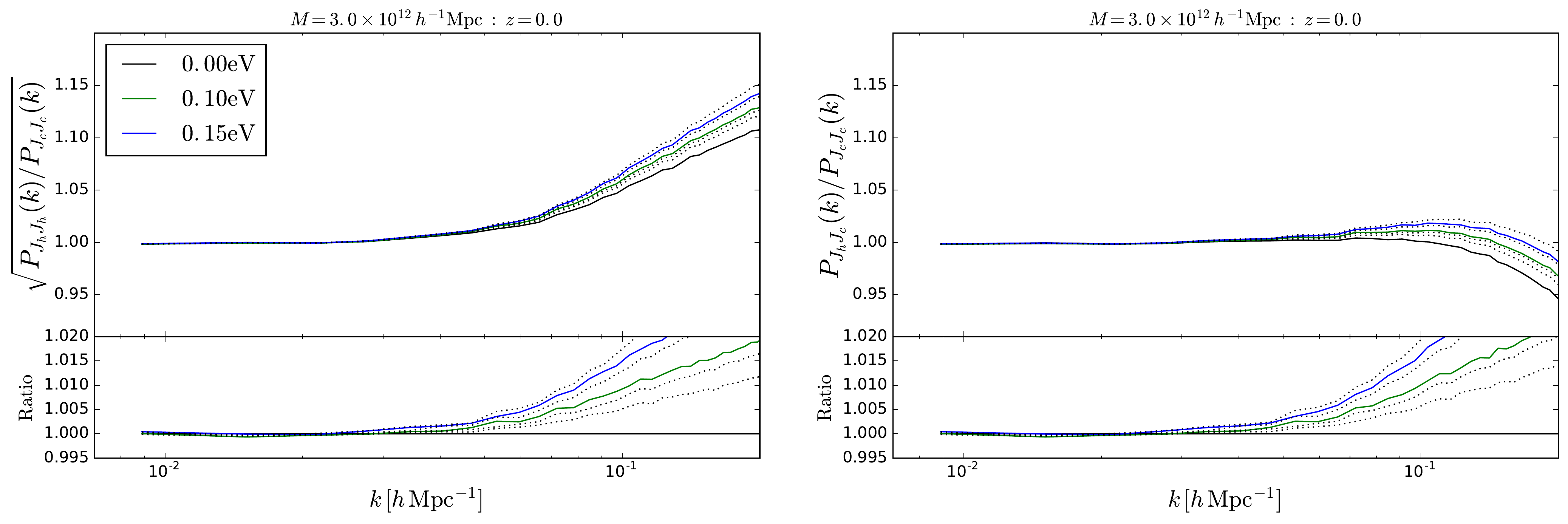}
\includegraphics[width=0.87\textwidth]{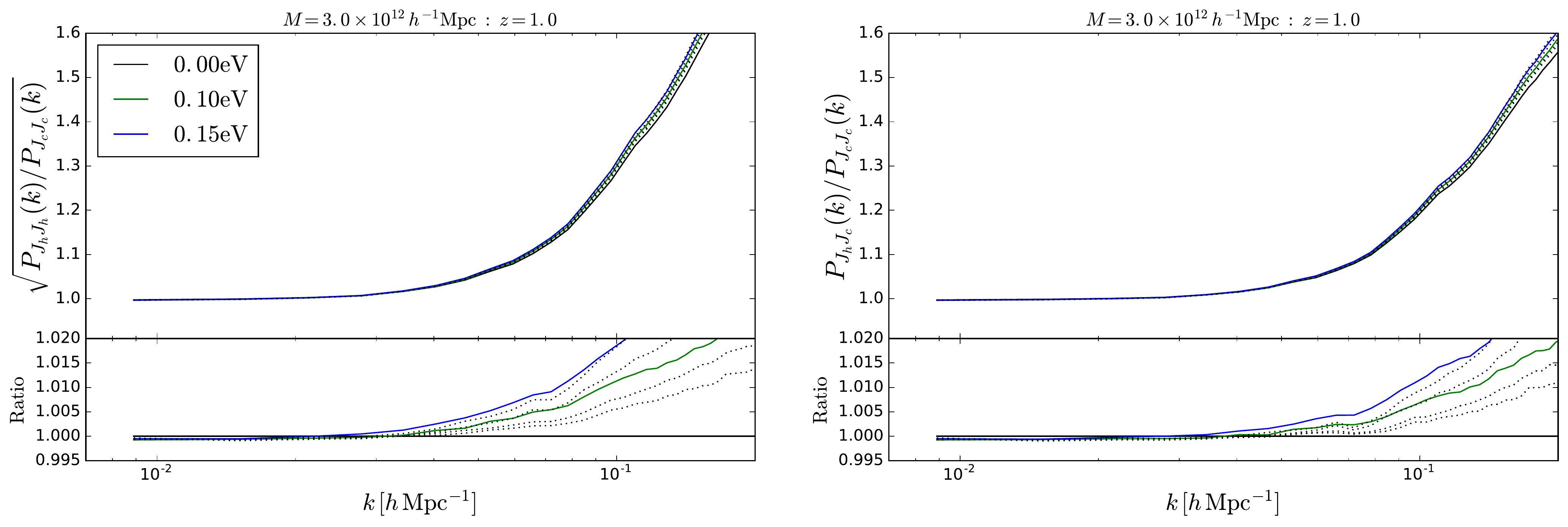}
\caption{Bias between the momentum of halos and the CDM+baryon fields. The momentum bias is estimated using the square root of the ratio between the auto-power spectra of the halos momentum and the CDM+baryons momentum (left column), and as the ratio between the cross-power spectrum of the halos and CDM+baryon momenta to the CDM+baryons momentum auto-power spectrum (right column). The first two rows show the results for halos with masses above $3.2\times10^{13}~h^{-1}M_\odot$ at $z=0$ (first row) and $z=0.5$ (second row). The third and fourth rows display the results for halos with masses above $3\times10^{12}~h^{-1}M_\odot$ at $z=0$ (third row) and $z=1$ (fourth row). Results are shown for models with massless neutrinos, fiducial (solid black) and models with lower $\sigma_8$ (dotted black, see Fig. \ref{fig:Pk_lin}), while models with 0.06 eV, 0.10 eV and 0.15 eV are shown in red, green and blue, respectively. This implies that in cosmologies with massive neutrinos the halo momentum do not trace the momentum of the underlying matter field. It is important to account for this in order to extract unbiased neutrino information from velocity/momentum observables, such as kSZ surveys.}
\label{fig:momentum_bias}
\end{center}
\end{figure*}

In order to avoid these problems we employ the momentum, $\vec{J}=(1+\delta)\vec{v}$, instead of the velocity field, since that is a quantity that is well defined everywhere. We then define the momentum bias either as
\be
b^2_J(k)=\frac{P_{J_{\rm h}J_{\rm h}}(k)}{P_{J_{\rm c}J_{\rm c}}(k)}
\ee
or as 
\be
b_J(k)=\frac{P_{J_{\rm h}J_{\rm c}}(k)}{P_{J_{\rm c}J_{\rm c}}(k)}~,
\ee
where $P_{J_{\rm h}J_{\rm h}}(k)$ and $P_{J_{\rm c}J_{\rm c}}(k)$ are the auto-power spectra of the halos and CDM+baryon momentum and $P_{J_{\rm h}J_{\rm c}}(k)$ is the cross-power spectrum between the momentum of halos and the momentum of CDM+baryons. We believe that on large scales, where the halos and CDM+baryon overdensities are small, the momentum bias should resemble the velocity bias. 

In Fig. \ref{fig:momentum_bias} we plot the momentum bias at different redshifts using the above two different definitions and for halos with masses above $3.2\times10^{13}~h^{-1}M_\odot$ and $3\times10^{12}~h^{-1}M_\odot$ for different models with massive and massless neutrinos. For the halos with masses above $3.2\times10^{13}~h^{-1}M_\odot$ we use the 100 realizations of the low-resolution simulations while for the halos with $M>3\times10^{12}~h^{-1}M_\odot$ we use the 13 high-resolution realizations. 

We find that the momentum bias on large scales, for all halo masses, redshifts and neutrino masses shown in Fig. \ref{fig:momentum_bias} tends to 1. The bottom part of each panel shows the relative difference between the momentum bias of each model with respect to the results of the fiducial model.  Our results point out that relative differences on large scales are completely negligible (below $0.5\%$) for all models. 

We thus conclude that we do not find evidence for a bias between the momentum of halos and the momentum of CDM+baryons. While the momentum receives contributions from the density even on large scales, the fact that the momentum bias is equal to 1 for halos of different mass, at different redshifts and for different cosmologies tends to indicate the lack of velocity bias between halos and CDM+baryons on large scales. If that would not be true, a very precise cancelation between the contribution from density and velocity would be require to produce such trend. The lack of momentum/velocity bias is in tension with the claim of \cite{Okoli_2017}, who predicted a velocity bias between CDM+baryons and halos on large-scales due to to the dynamical friction induced on halos by the cosmic neutrino background.

We remark that since the statistical properties of the velocity field of CDM+baryon will be different to the ones of the matter field for models with massive neutrinos, there will be a velocity/momentum bias between the halos and the underlying total matter field. It is important to account for this when extracting information from velocity/momentum observations, such as kSZ surveys, in order to avoid biases in the values of the derived parameters.

\begin{figure*}
\begin{center}
\includegraphics[width=1.0\textwidth]{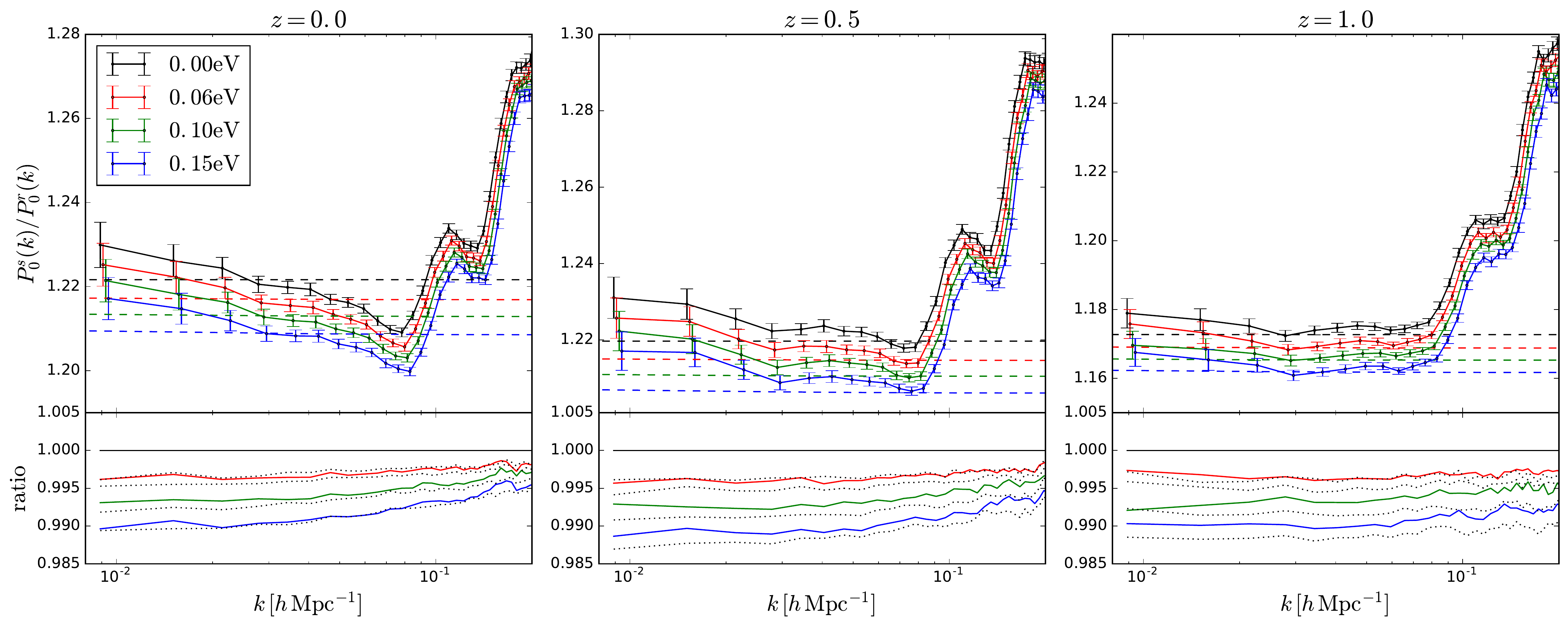}\\
\includegraphics[width=1.0\textwidth]{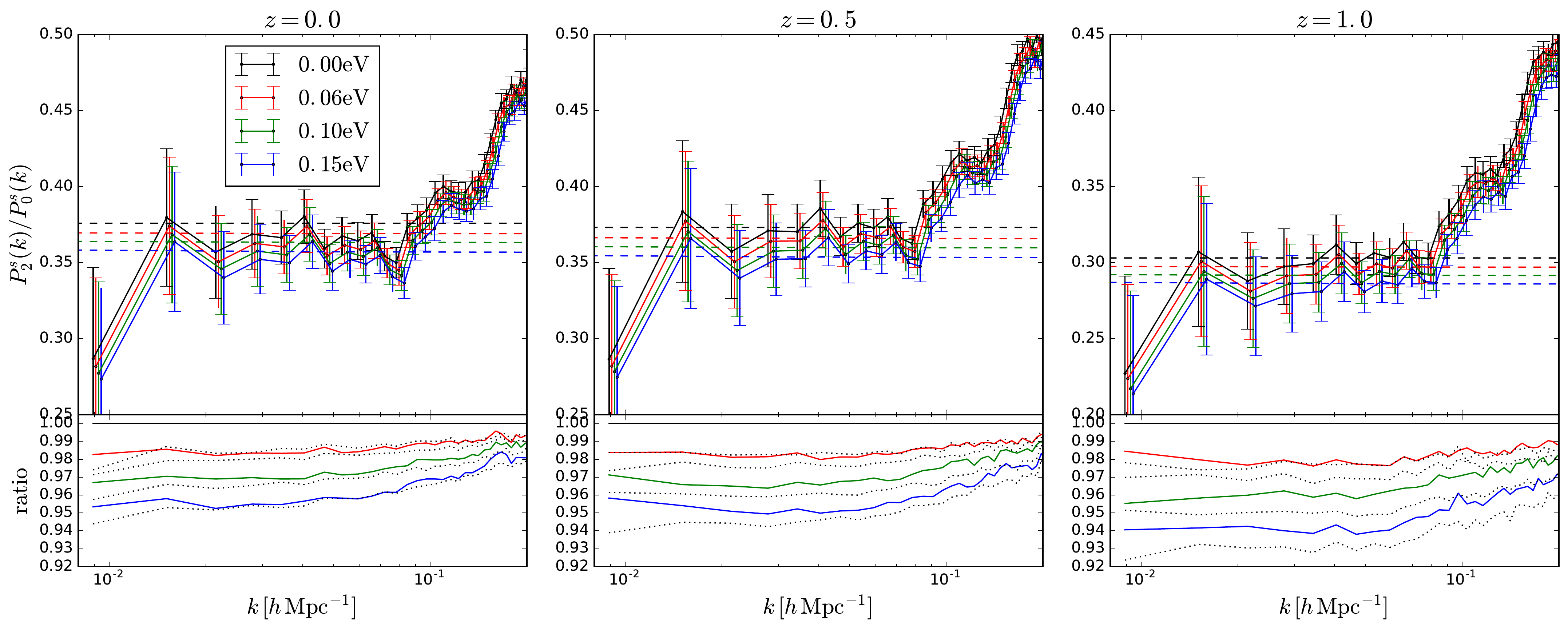}
\caption{We investigate the range of the scales where linear theory is able to describe the clustering of halos in redshift-space for models with massive and massless neutrinos. \textbf{Top:} Ratio between the halos monopole in redshift- and real-space at $z=0$ (left), $z=0.5$ (middle) and $z=1$ (right). We show results for models with 0.00 eV, 0.06 eV, 0.10 eV and 0.15 eV in solid black, red, green and blue lines, respectively. Error bars show the expected error from a survey covering a volume of $100~h^{-1}$Gpc. Linear theory predictions are shown with dashed lines. The bottom panels show the lines normalized by the results of the massless neutrino model. Dotted lines display the results for models with massless neutrinos and different values of $\sigma_8$. \textbf{Bottom:} same as the top panel but for the ratio between the halos quadrupole and monopole in redshift-space. As expected, non-linearities growth with time and linear theory validity shifts to larger scales as redshift decreases. Linear theory can reproduce the ratio between the halos quadrupole and monopole up to $k=0.07, 0.08, 0.09~h^{-1}{\rm Mpc}$ at $z=0$, 0.5 and 1, respectively. Similar conclusions can be derived for the monopoles ratio.}
\label{fig:halos_monopole}
\end{center}
\end{figure*}

\subsection{Comparison with linear theory}

We now investigate the scales where linear theory is able to describe the clustering properties of halos in redshift-space. Following the results of this paper and previous works such as \cite{Castorina_2015}, we can write the halos power spectrum in redshift-space, $P^s_{\rm hh}(k,\mu)$, as
\be
P^s_{\rm hh}(k,\mu)=(b+f_{\rm cb}\mu^2)^2P_{\rm cb}(k)~,
\ee
with $b$ being the halo bias and $f_{\rm cb}$ and $P_{\rm cb}(k)$ are the growth rate and power spectrum of CDM+baryons. 

The justification of the above expression is as follows: the halos clustering resembles the clustering of the CDM+baryons field, instead of the total matter field \citep{Villaescusa-Navarro_2014, Castorina_2014}, and therefore the last term in the r.h.s. of the above equation should be the CDM+baryons power spectrum. Our results suggest that halo velocities follow the velocities of the CDM+baryons field on large scales, so no velocity bias should be added to the above equation. Finally, given the fact that halo properties are controlled by the CDM+baryon field it is expected that the growth rate in the above equation will be the one of CDM+baryons instead of total matter \citep[see][for a similar justification]{Castorina_2015}.

We show the ratio between the halos monopole in redshift- and real-space at different redshifts in the top panels of Fig. \ref{fig:halos_monopole}. The amplitude of that ratio, on large scales, is given by $1+2\beta/3+\beta^2/5$ in linear theory, with $\beta=f_{\rm cb}/b$, that we show in that figure with dashed lines. The errorbars on the ratios represent the standard deviation of the mean from the 100 realizations, and therefore the expected error for a volume of $100~(h^{-1}{\rm Gpc})^3$. The halo linear bias is estimated by fitting the ratio between the halo-CDM cross-power spectrum over the CDM auto-power spectrum to a constant line over the scales $k\leqslant0.05~h^{-1}{\rm Mpc}$. The linear growth rate is obtained from CAMB. 

We find that, given the errorbars of our measurements, linear theory is not particularly good in describing the amplitude of our results on large scales. As expected, the agreement with linear theory increases with redshift. At $z=0$, even on the largest scales we probe in our simulations, we find a scale-dependence in the monopoles ratio that can be not described by linear theory. Notice however that we find an overall amplitude offset between our results and predictions by linear theory at $z=0.5$, although it is not significant: $\sim1\sigma-1.5\sigma$. This can be due to an underestimation of the value of the linear bias from our fits.

We show the monopoles ratios, normalized by the results of the fiducial model, in the lower part of the top panels of Fig. \ref{fig:halos_monopole}. We see that the effect of massive neutrinos with masses of 0.15 eV is limited to $1\%$. We find that models with massive neutrinos change the monopoles ratio in a scale-dependent way, with the amplitude on large scales more suppressed than on small scales, with respect to the fiducial model. This effect is mainly driven by $\sigma_8$, as we find from the dotted lines, showing the results for models with massless neutrinos and lower values of $\sigma_8$. This points out that it is the change in the bias, mainly controlled by $\sigma_8$, and not in the growth rate, what drives the differences.

The bottom panels of Fig. \ref{fig:halos_monopole} show the ratio between the halos quadrupole and monopole in redshift-space at several redshifts. Linear theory predicts an amplitude on large scales for that ratio equal to $(4\beta/3+4\beta^2/7)/(1+2\beta/3+\beta^2/5)$, that we show with dashed lines for the different models. We find good agreement between our results and linear theory on large scales, that improves with redshift on small scales. This is however likely due to the fact that errorbars on this quantity are still pretty large. The error bars on this ratio are however larger than those from the monopoles ratio, since cosmic variance affects more strongly the quadrupole than the monopole. 

In the lower part of those panels we display the results normalized by the output of the fiducial model. As in the case of the monopoles ratio, we find that massive neutrinos induce an amplitude reduction that it is scale dependent. That effect is however mainly driven by $\sigma_8$, since we find a strong degeneracy between the results of the massive neutrinos and the models with massless neutrinos and lower $\sigma_8$ (dotted lines). 

We conclude that measuring neutrino masses on linear scales with redshift-space distortions is challenging, even with unrealistic volumes as large as $100~(h^{-1}{\rm Gpc})^3$. Non-linear corrections are important even on very large-scales and for multipoles ratios. Redshift-space distortions can be used as complementary source of cosmological information to standard methods to weigh neutrinos as those using the shape of the galaxy power spectrum.

\section{Baryonic effects}
\label{sec:baryons}

Here we investigate, for the first time, the impact of baryonic effects on redshift-space distortions in cosmologies with massive neutrinos.

\begin{figure*}
\begin{center}
\includegraphics[width=1.0\textwidth]{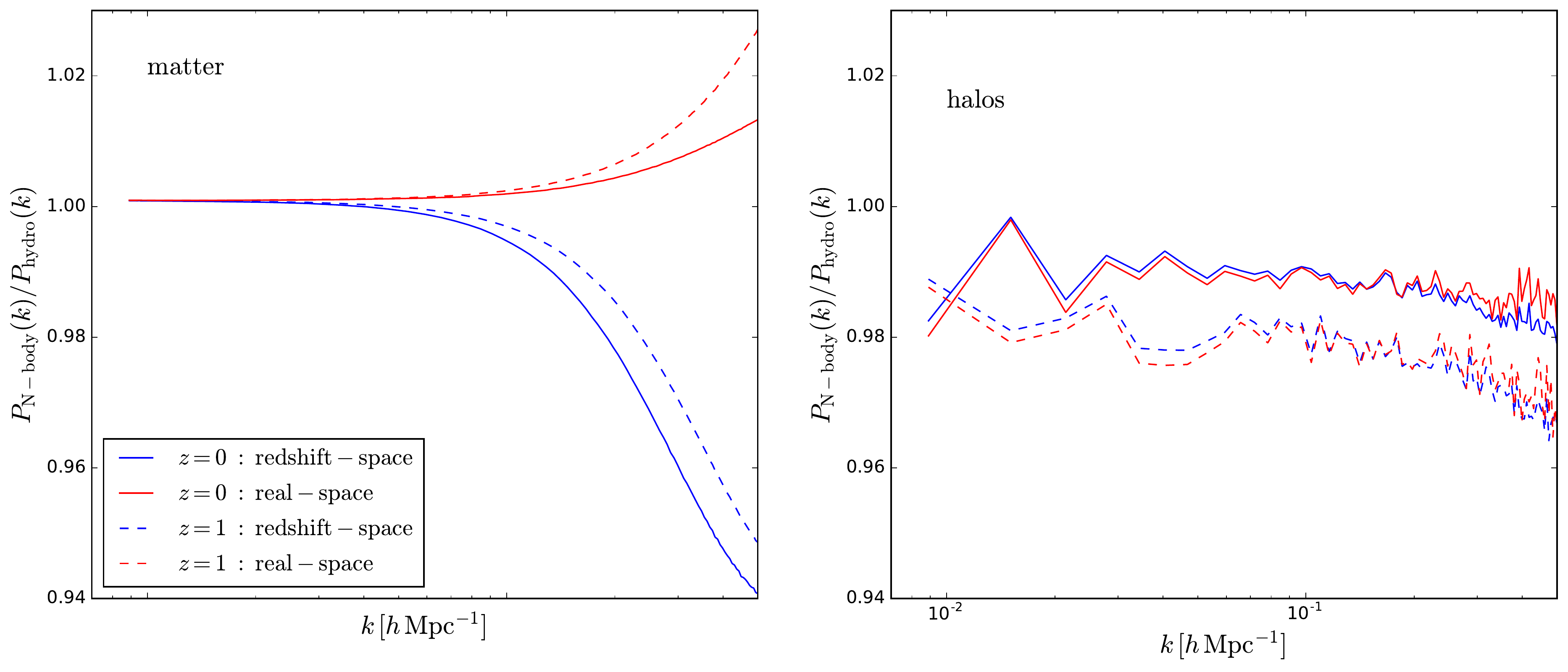}
\caption{Impact of baryons on the clustering of matter (left) and halos (right) in real- (red) and redshift-space (blue) at $z=0$ (solid) and $z=1$ (dashed). The plot shows the ratio between the power spectrum from the N-body simulations to the hydrodynamics simulations. Baryons affect the shape and amplitude of the matter power spectrum in redshift-space by up to $5\%$ at $k=0.5~h^{-1}$Mpc while for halos the effect is mainly an overall amplitude suppression of $\simeq1.5\%$.}
\label{fig:hydro_abs}
\end{center}
\end{figure*}
 
It is well known that baryonic effects imprint their signature on the spatial distribution of matter \citep[e.g.][]{Van_Daalen_2011}, affecting both the shape and amplitude of the matter power spectrum. This represents a major complication for cosmology, as the physics and the efficiency involved in those astrophysical processes is not fully understood. We will however show that the relative effect induced by massive neutrinos can be, to a very good approximation ($1\%$ down to $k=0.5~h{\rm Mpc}^{-1}$), factorized out. This means that neutrino effects can be written as a transfer function that can be calibrated using N-body instead of fully hydrodynamic simulations. We notice that \cite{Mummery} carried out a similar analysis in real-space, reaching similar conclusions to ours. 

In Fig. \ref{fig:hydro_abs} we show the effect of baryons on the absolute amplitude and shape of the power spectrum of matter and halos in real- and redshift-space at redshifts $z=0$ and $z=1$. In our simulations we find that the effects of baryons on the matter power spectrum in real-space is limited to $1-2\%$ at those redshifts, in agreement with previous works \citep[e.g.][]{Vogelsberger_2014}. We notice however that this is also caused by the way we set the smoothing lengths of the gas particles (see appendix \ref{sec:sph} for a detailed analysis). On the other hand, in redshift-space, baryonic effects can be as large at $6\%$ at $z=0$. We notice that previous works \citep{Hellwing_2016} have found the magnitude this effect to be slightly lower than our findings. We have verified that our results are robust against the way we set the smoothing lengths of the gas particles (see appendix \ref{sec:sph}). It is important to emphasize that the changes induced by baryons on the CDM+baryon field are as large as those caused by neutrinos with 0.15 eV masses (see Fig. \ref{fig:matter}). 

The effect of hydrodynamics and astrophysical processes on the clustering of halos has a magnitude of $1-3\%$, with a very mild dependence on scale down to $0.5~h{\rm Mpc}^{-1}$ that correlates with redshift.

We study the relative differences in the monopoles, quadrupoles and fully 2D power spectrum of the matter, CDM+baryon and halos fields induced by massive neutrinos when employing N-body and hydrodynamic simulations. Given the fact that hydrodynamic simulations are much more computationally expensive than N-body simulations, we focus our analysis on just one model with 0.15 eV massive neutrinos. 

\begin{figure*}
\begin{center}
\includegraphics[width=1.0\textwidth]{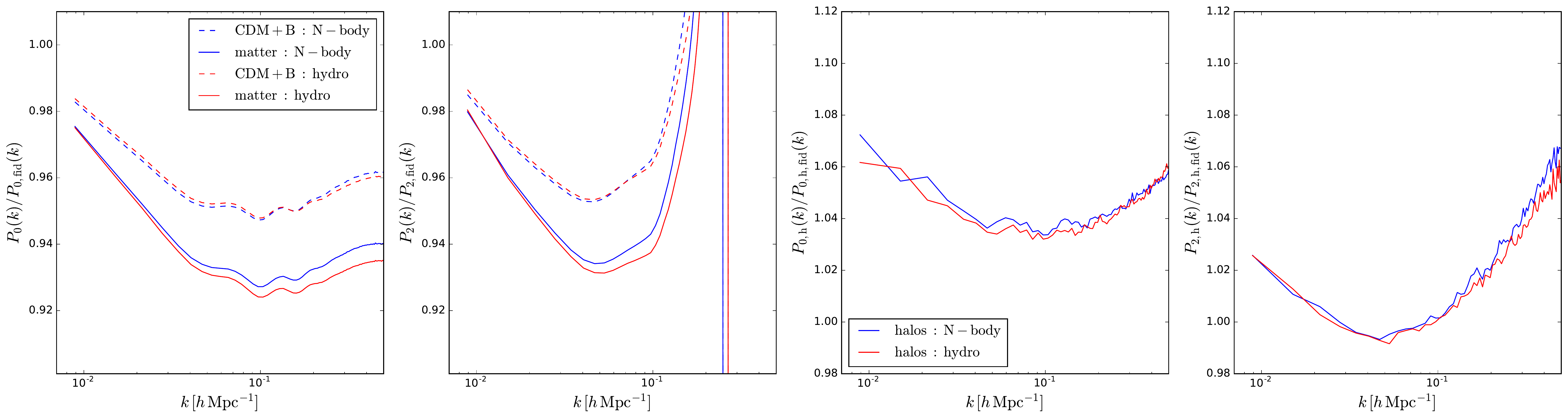}\\
\includegraphics[width=1.0\textwidth]{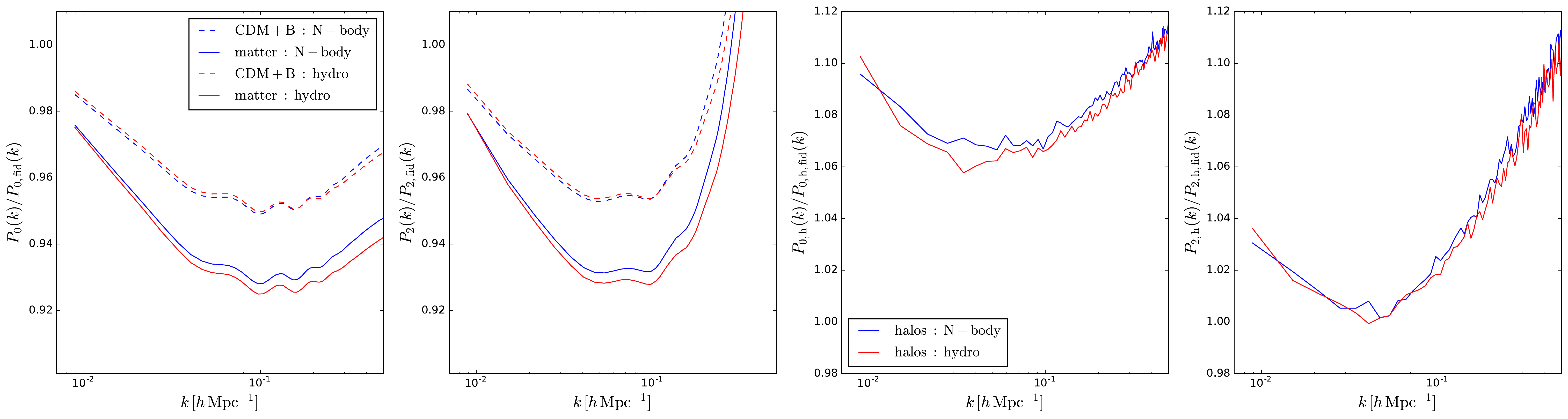}\\
\includegraphics[width=0.49\textwidth]{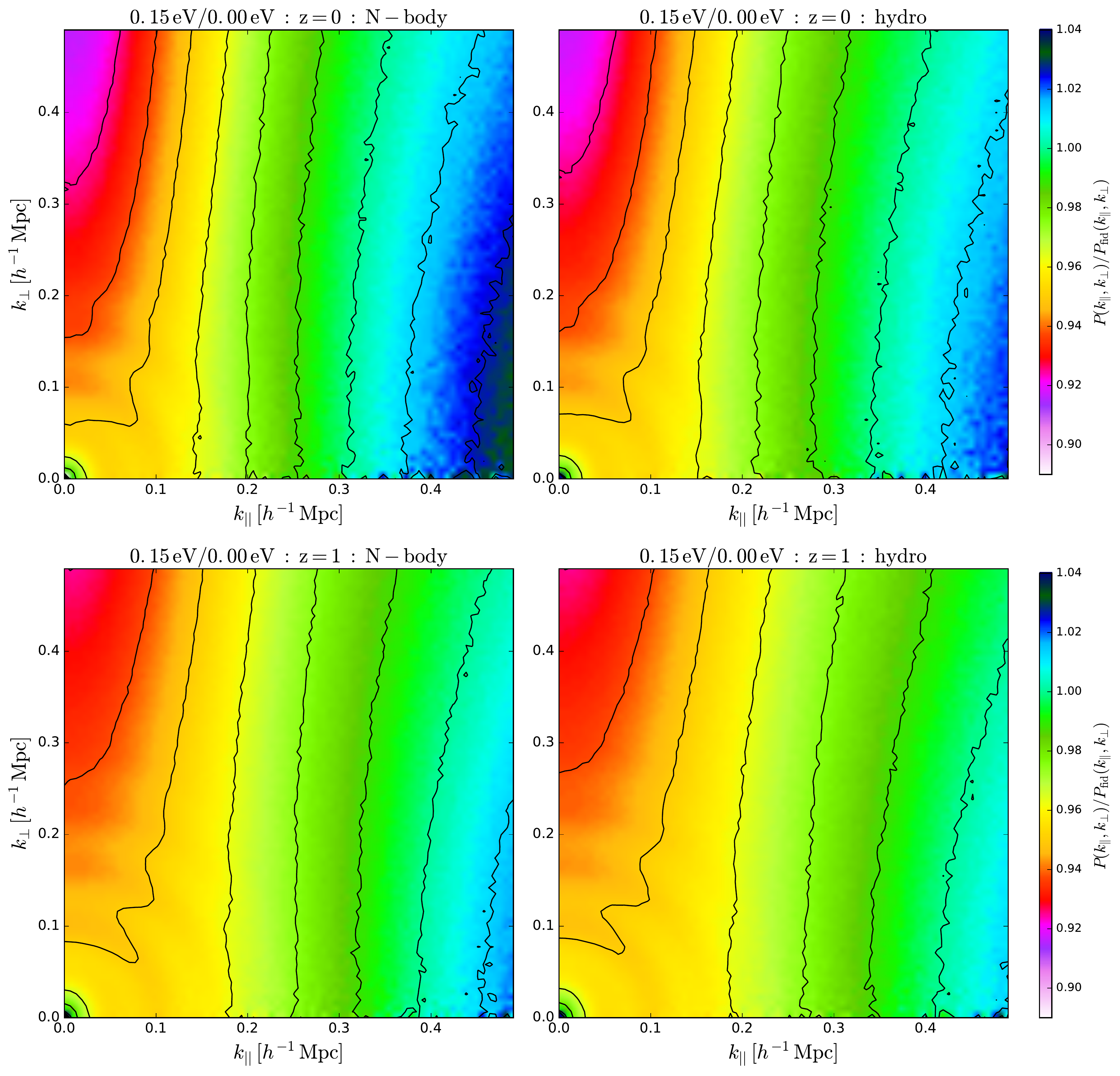}
\includegraphics[width=0.49\textwidth]{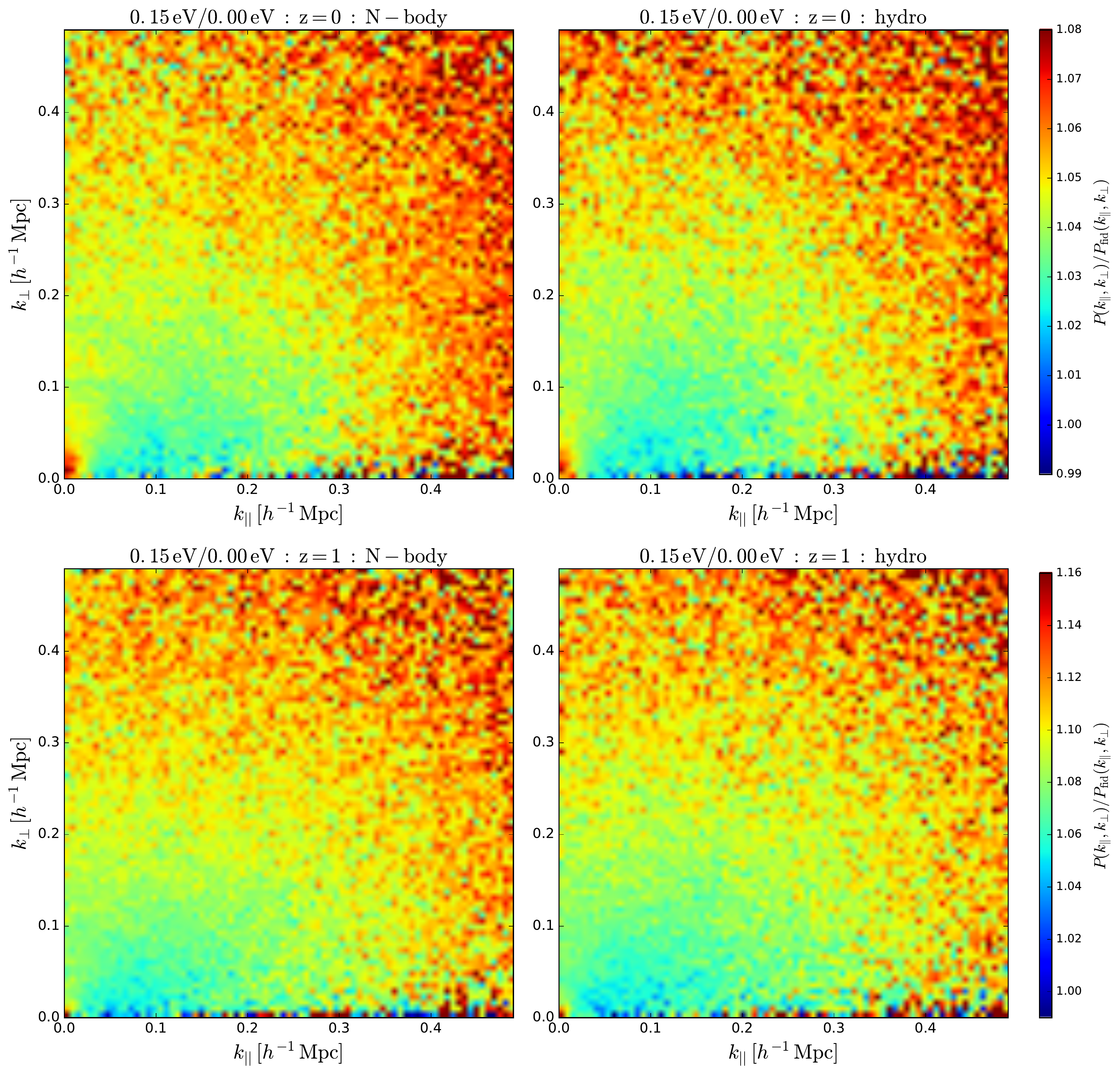}
\caption{Baryonic effects on redshift-space distortions on the matter, CDM+baryons and halo fields for a model with 0.15 eV neutrinos. \textbf{Top:} Monopoles (first column) and quadrupoles (second column) of CDM+baryons and matter fields and the monopoles (third column) and quadrupoles (fourth column), normalized by the results of the fiducial model, at $z=0$ (upper row) and $z=1$ (bottom row). Results are shown for the N-body and hydrodynamic simulations (see legend).  \textbf{Bottom:} 2D power spectrum of the CDM+baryon field (left panels) and the halo field (right panels), normalized by the results of the fiducial cosmology. We show the results at $z=0$ and $z=1$ and for the N-body and hydrodynamic simulations (see top part of each panel). Black lines show iso-power contours. The relative effects that neutrinos induce barely change if baryons and their effects are taken into account. Thus, neutrino effects can be written as a transfer function that, within $1\%$ down to $k=0.5~h{\rm Mpc}^{-1}$, do not depend on baryons. That transfer function can studied through N-body simulations rather than hydrodynamical simulations.}
\label{fig:hydro}
\end{center}
\end{figure*}

We show on the first column of the top part of Fig. \ref{fig:hydro} the monopoles of CDM+baryons and matter for the 0.15 eV model in redshift-pace at $z=0$ and $z=1$, normalized by the results of the fiducial model, when using N-body and hydrodynamic simulations. As can be seen, for CDM+baryons, the relative difference is barely affected by hydrodynamics (well below $1\%$) down to $k=0.5~h{\rm Mpc}^{-1}$ at both redshifts. We find that baryons induce slightly larger differences in the matter field, but still below $1\%$. The second column of the top part displays the results for the quadrupole of the CDM+baryons and matter fields. Also in this case we find that hydrodynamics and astrophysical effects only change the relative difference caused by neutrinos by less than $1\%$. 

The third and fourth columns of the top part of Fig. \ref{fig:hydro} show the results for the halos monopole and quadrupole in redshift-space, respectively. As for the matter and CDM+baryon fields, we find that baryons do not change appreciably the relative effect induced by neutrino masses. Given the error bars arising from cosmic variance, we conclude that the results from N-body and hydrodynamic simulations are in very good agreement.

Finally, we investigate whether baryons affect the morphology of the relative difference induced by massive neutrinos on the CDM+baryon and halo fields. We show the results in the bottom part of Fig. \ref{fig:hydro} for the CDM+baryon field (left panels) and the halo field (right panels) at redshifts $z=0$ (upper row) and $z=1$ (bottom row). We find that the angle-dependence of the relative effect of massive neutrinos is not much affected by baryons, at both redshifts and for both halos and CDM+baryons. Our results point out that while the relative difference in the 2D power spectrum of CDM+baryons is almost unaffected by baryons for modes with low $k_\|$ (independently of $k_\bot$), we observe a larger effect for modes with large $k_\|$, although differences are below $1\%$. This effect is however averaged out when estimating the relative difference in the monopole, as can be seen in the upper part of Fig. \ref{fig:hydro}. This happens because only a relatively small fraction of modes are affected by this effect.

We reach similar conclusions by looking at the halo field, where baryons barely affect the 2D structure of the relative difference between 0.15eV and massless neutrinos. Unfortunately, cosmic variance is still large enough so that we can not quantify any baryonic effect on the amplitude or angle dependence of the ratio. We have verified that our results are robust against the way we set the smoothing lengths of the gas particles (see appendix \ref{sec:sph}).

We thus conclude that even though baryons can largely affect the absolute amplitude and shape of the clustering pattern of matter, CDM+baryons and halos in redshift-space, massive neutrinos effects can be factorized out and relative differences are robust to hydrodynamics and astrophysical processes within $1\%$, for the observables we have considered in this paper down to $k=0.5~h{\rm Mpc}^{-1}$. We emphasize that a study like the one presented here, where the fully 2D power spectrum in redshift-space is compared by varying cosmological models and baryon effects, can only be carried out given the very large volumes we have accessible through our numerical simulations.
\vspace{0.5cm}

\section{Conclusions}
\label{sec:conclusions}

Measuring the sum of the neutrino masses is one of the most important challenges in modern cosmology. Upcoming large scale structure surveys are expected to have enough statistical power to be able to detect the minimum sum of the neutrino mass allowed by neutrino oscillation experiments.

Unfortunately, our ability to extract cosmological information from galaxy clustering measurements is limited by several physical processes: 1) the density field becomes non-linear on small scales and/or low redshifts due to gravitational evolution, 2) we do not observe the density field directly, but only tracers of it such as galaxies or cosmic neutral hydrogen, 3) the redshifts of the tracers we measure are affected by their peculiar velocities, 4) astrophysical processes such as feedback may potentially affect the distribution of matter and peculiar velocities in a non-trivial way. 

Precise theory predictions are required to measure the sum of the neutrino masses robustly from redshift surveys. Since the effect of neutrino masses is so small, any inaccuracy in the theory side can give rise to biases in both the sum of the neutrino masses and the value of the other cosmological observables. For neutrinos, these biases can incorrectly lead to major consequences. For instance, a cosmological detection of neutrinos masses below $\simeq0.06$ eV at $3\sigma$ will have major implications for both cosmology and particle physics. 

It is thus timely to understand and model as accurate as possible the impact of non-linearities, halo/galaxy bias, redshift-space distortions and baryonic effects on cosmologies with massive neutrinos. While the above first three complications have been carefully studied in cosmologies with massless neutrinos (and only partially the fourth), only the first and the second have been systematically studied for models with massive neutrinos. 

In this work we have analyzed in detail the following aspects: 1) the impact of neutrinos and $\sigma_8$ on the clustering of matter, CDM+baryons and halos in redshift-space in the fully non-linear regime, 2) the impact of neutrinos on the non-linear growth rate, 3) the bias between the momentum of halos and the CDM+baryon field, 4) the scales where linear theory is able to describe the clustering properties of halos in models with massive neutrinos and 5) the impact of baryons on redshift-space distortions in cosmologies with massive and massless neutrinos. 

We investigate the above points using a very large set containing more than 1000 N-body and hydrodynamical simulations with realistic neutrino masses. For each model we have 100 low-resolution N-body realizations, where we resolve halos with masses above $\simeq3\times10^{13}~h^{-1}M_\odot$ in a box of 1 $h^{-1}{\rm Gpc}$, and 13 high-resolution N-body realizations, resolving halos with masses above $\simeq3\times10^{12}~h^{-1}M_\odot$ in a box of 1 $h^{-1}{\rm Gpc}$. We simulate three different cosmologies with 0.06 eV, 0.10 eV and 0.15 eV degenerate massive neutrinos. We also have 5 different cosmologies with massless neutrinos and different values of $\sigma_8$ that more or less match the value of either $\sigma_8^c$ or $\sigma_8^m$ from the models with massive neutrinos. For the fiducial and $M_\nu=0.15$ eV models we also have 100 low-resolution hydrodynamic simulations.

We now summarize the main conclusions of this work:

\begin{itemize}

\item We find that massive neutrinos and $\sigma_8$ produce a very different effects on the clustering of matter, CDM+baryons and halos in redshift-space on large scales. On the other hand, they produce almost identical effects on the clustering of CDM+baryons on small scales. That degeneracy is also present in the matter field, but can be partially broken through the quadrupole. The magnitude of this degeneracy is smaller for the clustering of halos, although both parameters produce, within a few percent, almost the same effect. 

In the near future, theory predictions for galaxy clustering in redshift-space may arise by combining the output of N-body simulations with halo occupation distribution (HOD) models. Given the degeneracy between $M_\nu$ and $\sigma_8$ on small scales, the neutrino signal may be partially mimicked by assembly bias, since both assembly bias and massive neutrinos can change the amplitude of the clustering on large scales keeping fixed the small scale clustering.

\item The amplitude of the growth rate of the CDM+baryon field in cosmologies with massive neutrinos is lower, in scales $k\simeq10^{-1}-0.5~h{\rm Mpc}^{-1}$, than the one of the corresponding massless neutrino cosmology. That suppression is however very small: less than $2\%$ for 0.15 eV neutrinos at $k=0.5~h{\rm Mpc}^{-1}$. On small scales, $\sigma_8$ and $M_\nu$ produce very similar effects. 

\item We find no bias between the momentum of halos and CDM+baryon on large scales. Since in cosmologies with massive neutrinos the properties of the CDM+baryons and matter fields are different, this implies that there is a momentum bias between halos and matter. It is important to account for this when extracting neutrino information from velocity/momentum observations as kSZ surveys.

\item Our results indicate that, at linear order and in cosmologies with massive neutrinos, the galaxy power spectrum in redshift-space, $P_{\rm gg}^s(k)$, can be written as
\be
P^s_{\rm gg}(k)=\left(b+f_{\rm cb}(k)\mu^2\right)^2P_{\rm cb}(k)~.
\ee
where $b$ is the galaxy bias and $f_{\rm cb}$ and $P_{\rm cb}(k)$ are growth rate and power spectrum of the CDM+baryons field, respectively. We find that the above expression can reproduce the clustering properties of dark matter halos down to $k=0.07, 0.08, 0.09~h^{-1}{\rm Mpc}$ at $z=0$, 0.5 and 1, respectively.

\item We show that baryonic effects can affect the shape and amplitude of the CDM+baryons and matter power spectrum by as much as $\simeq5\%$ at $k=0.5~h{\rm Mpc}^{-1}$ at $z=0$. The effect on halos is smaller: $\simeq 1\%$ and it has a weak dependence with scale. We show that the relative effect that neutrino induce on the clustering of matter, CDM+baryons and halos, i.e. $P_{M_\nu}(k,\mu)/P_{M_\nu=0}(k,\mu)$, is affected by baryons by less than $1\%$ down to $k=0.5~h{\rm Mpc}^{-1}$. This means that the neutrino effects (at least at to $k=0.5~h{\rm Mpc}^{-1}$) can be written as a transfer function that, to a very good approximation, do not depend on baryons. That transfer function can be calibrated using N-body instead of hydrodynamical simulations.

\end{itemize}

This work represents a step forward into a careful and systematic description of neutrino effects on cosmological observables in the mildly non-linear regime.

\section*{ACKNOWLEDGEMENTS} 
Numerical simulations have been run in the Flatiron institute using the \textit{rusty} cluster. We thank Bhuvnesh Jain and Ravi Sheth for useful conversations. FVN thanks Matteo Zennaro for his help setting up the initial conditions of the numerical simulations. The work of FVN and DNS is supported by the Simons Foundation.

\begin{appendix}

\section{Halo bias}
\label{sec:halo_bias}

Here we show the value of the halo linear bias we measure from our simulations. We measure the halo bias as
\be
b_{\rm h}(k)=\frac{P_{\rm ch}(k)}{P_{\rm cb}(k)}
\ee
where $P_{\rm ch}(k)$ is the CDM+baryon-halos cross-power spectrum and $P_{\rm cb}(k)$ is the CDM+baryons auto-power spectrum, both in real-space. For each model, we measure the halo bias using the above equation for each of the 100 realizations available. We then compute the mean and standard deviation and extract the value of the linear bias by fitting the results to a constant, $b_1$, over the range of the scales $k\leqslant5\times10^{-2}~h{\rm Mpc}^{-1}$.

In table \ref{tab:bias} we show the values of $b_1$, for halos with masses above $3.2\times10^{13}~h^{-1}M_\odot$, for the different cosmologies explored in this work at several redshifts.

\begin{table}
\caption{\label{tab:bias}Values of linear halo bias, for halos with masses above $3.2\times10^{13}~h^{-1}M_\odot$, for the different models at redshifts 0, 0.5, 1 and 2.}
\begin{center}
{\renewcommand{\arraystretch}{1.3}
\begin{tabular}{| c | c | c | c | c | c | c | c |}
\hline
Name & $M_\nu$ & $\sigma_8^m$ & $\sigma_8^c$ & \multicolumn{4}{|c|}{bias}\\
& & & &  $z=0$ & $z=0.5$ & $z=1$ & $z=2$\\
\hline
L0 (fid) & 0.00 & \multicolumn{2}{|c|}{0.833} & 1.739 & 2.524 & 3.633 & 6.839 \\
\hline
L0-1 & 0.00 & \multicolumn{2}{|c|}{0.822} & 1.770 & 2.573 & 3.712 & 7.000\\
\hline
L0-2 & 0.00 & \multicolumn{2}{|c|}{0.818} & 1.782 & 2.594 & 3.742 & 7.082\\
\hline
L0-3 & 0.00 & \multicolumn{2}{|c|}{0.807} & 1.814 & 2.643 & 3.821 & 7.255\\
\hline
L0-4 & 0.00 & \multicolumn{2}{|c|}{0.798} & 1.842 & 2.689 & 3.895 & 7.394\\
\hline
L6 & 0.06 & 0.819 & 0.822 & 1.766 & 2.566 & 3.690 & 6.942\\
\hline
L10 & 0.10 & 0.809 & 0.815 & 1.797 & 2.614 & 3.770 & 7.110\\
\hline
L15 & 0.15 & 0.798 & 0.806 & 1.828 & 2.660 & 3.840 & 7.258 \\
\hline
\end{tabular}}
\end{center}
\end{table}

\section{Effects from gas smoothing length}
\label{sec:sph}

The hydrodynamic simulations employed in this work has been run by setting the smoothing length of the gas particles equal to their SPH radii. The reason is to avoid the short-scale forces that produces an undesired coupling between CDM and gas \citep{Angulo_2013}. That coupling induces a significant deviation of the gas power spectrum on all scales, that translates into a shift with respect to linear theory on large scales. 

Unfortunately, some side effects arise by doing that, such as gas losing power with respect to linear theory at high redshift on small scales. In order to test the robustness of our results against these effects, we have run an equivalent set of hydrodynamical simulations to those in table \ref{tab:i} by setting the smoothing length of the gas particles to $1/40$ their mean inter-particle distance, i.e. the same as for CDM. We have then repeated the analysis of section \ref{sec:baryons}.

\begin{figure}
\begin{center}
\includegraphics[width=1.0\textwidth]{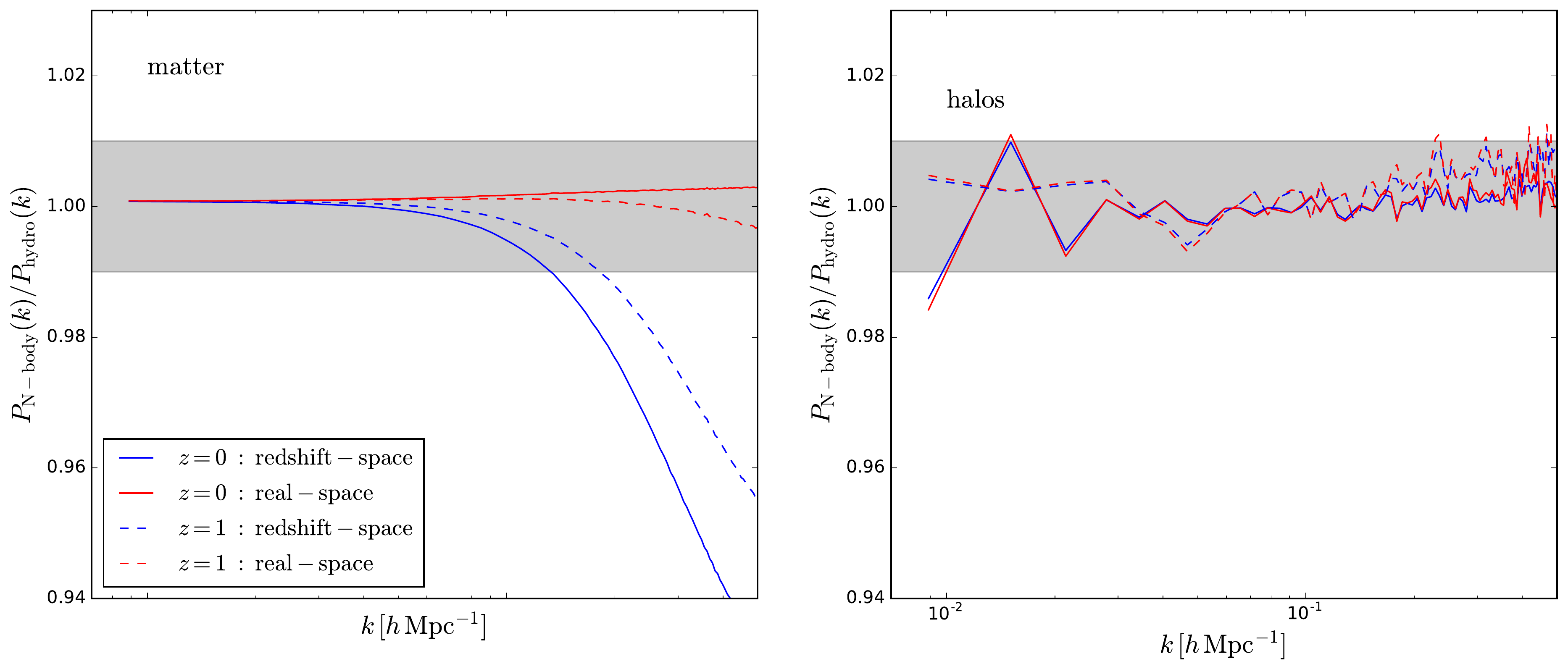}
\caption{Same as Fig. \ref{fig:hydro_abs} when the smoothing lengths of the gas particles are set to $1/40$ of their mean inter-particle distance. In our standard hydrodynamic simulations we set the smoothing lengths of the gas particles to be equal to their SPH radii.}
\label{fig:hydro_abs_new}
\end{center}
\end{figure}

The effect of baryons on the matter and halo power spectrum in real- and redshift-space is shown in Fig. \ref{fig:hydro_abs_new}. We find that baryonic effects in real-space are now below the percent level at both $z=0$ and $z=1$, in agreement with \cite{Hellwing_2016,Springel_2017}. This is due to the fact that the gas in our standard runs loose power at high-redshift, inducing a transient that still persists at low redshift. We notice that a way to, at least partially, avoid this is simply to increase the resolution of our standard runs. In our case, given the volumes we are interested in, this is computationally unfeasible. 

In redshift-space the matter clustering seems to be less sensitive to the way gas smoothing lengths are set. As with our standard runs, we find that baryons increase the amplitude of the monopole on small scales with respect to the N-body simulations. Our results are in tension with those of \citep{Hellwing_2016}, where the authors found a much smaller effect. Unfortunately, we can not discern whether this discrepancy is due to the way we generate the initial conditions of it is just an artifact of the relatively poor resolution of our hydrodynamic runs.  

The effect of the gas smoothing length on the halo clustering is mostly attributed to an overall change in the amplitude. The reason is that in our standard runs we have a deficit of power on small scales due to gas losing power on small scales at high-redshift. This manifests in an lower abundance of halos in those hydrodynamic runs than those in N-body simulations or hydrodynamic simulations with standard smoothing lengths. That lower abundance induce a higher amplitude of the halo clustering in our standard runs. By using the standard smoothing lengths for the gas particles we find that the effect of baryons on halo clustering in below $1\%$ down to $k=0.5~h{\rm Mpc}^{-1}$.

\begin{figure*}
\begin{center}
\includegraphics[width=0.49\textwidth]{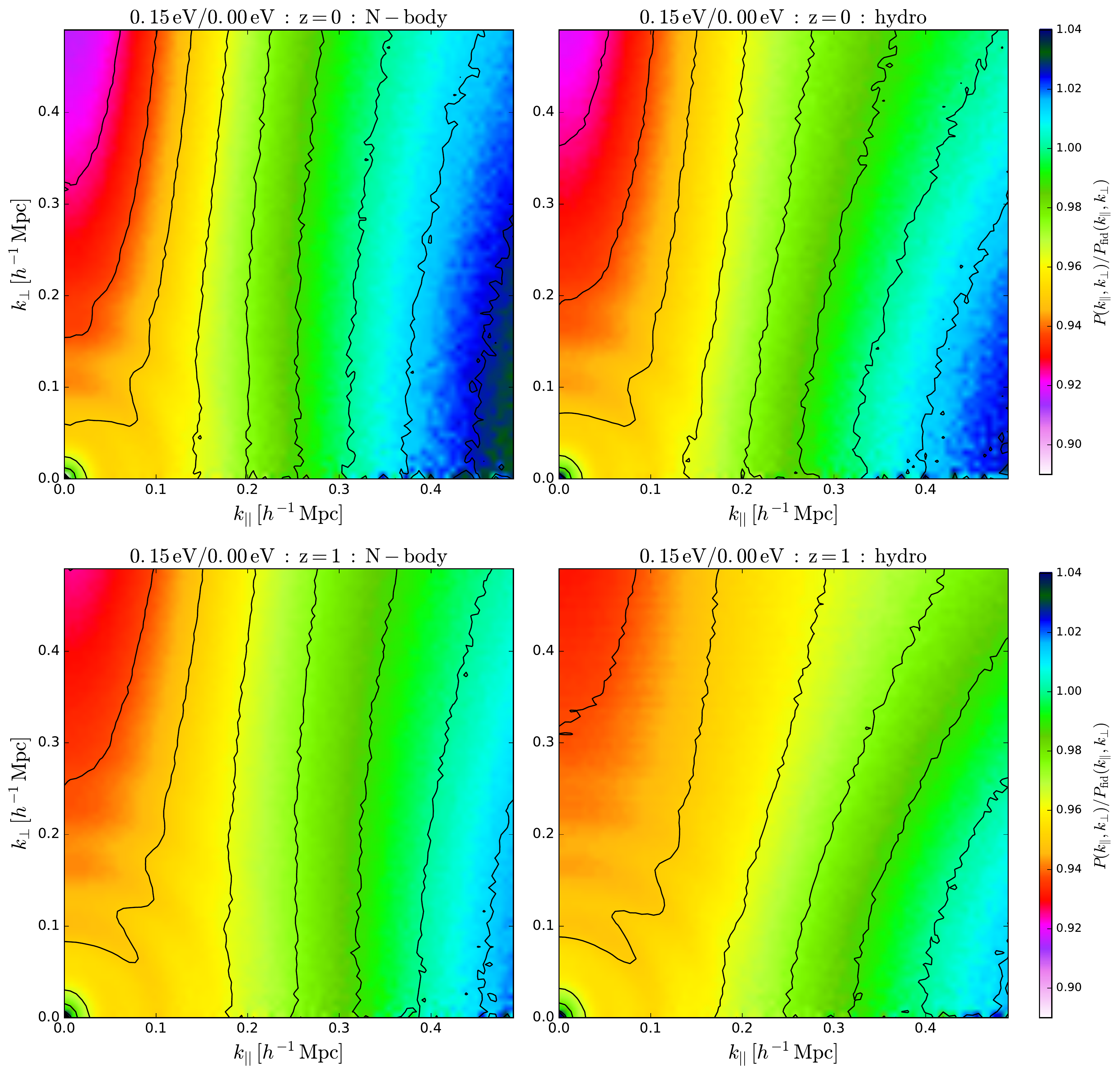}
\includegraphics[width=0.49\textwidth]{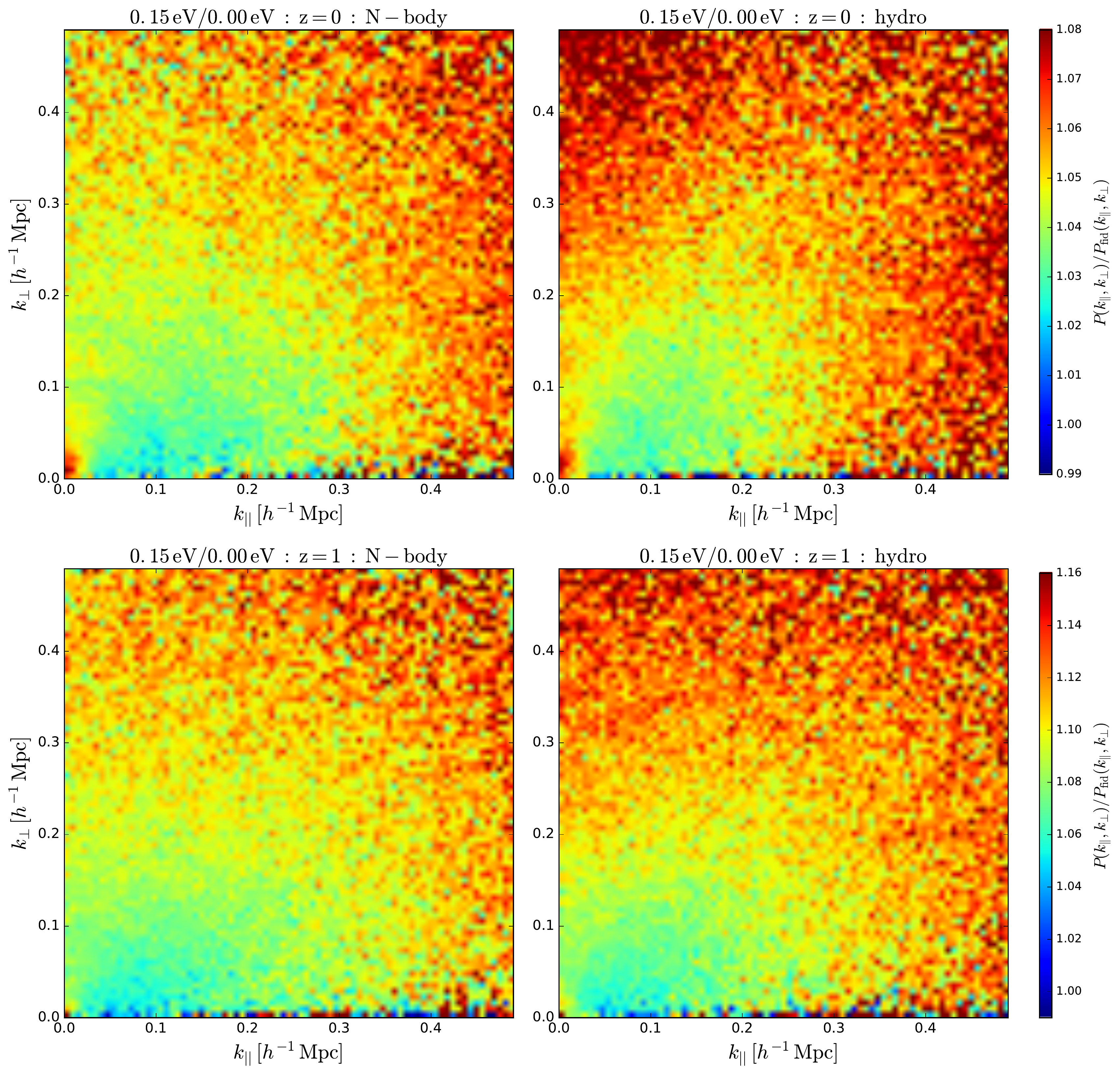}
\caption{Same as Fig. \ref{fig:hydro} when the smoothing lengths of the gas particles are set to $1/40$ of their mean inter-particle distance. In our standard hydrodynamic simulations we set the smoothing lengths of the gas particles to be equal to their SPH radii.}
\label{fig:hydro_new}
\end{center}
\end{figure*}

We finally check the impact of the gas smoothing lengths on our claim that neutrino effects can be factorized out and studied using N-body simulations. In Fig. \ref{fig:hydro_new} we show the equivalent of Fig. \ref{fig:hydro} when the smoothing lengths of the gas particles are set to $1/40$ of their mean inter-particle distance. We find that baryons imprint a larger modification on the relative difference that neutrino induce, with respect to our fiducial runs, for both CDM+baryons and halos. The effect is however of the order of $1\%$ and most of the differences take place on small scales. 

We thus conclude that by setting the smoothing lengths of the gas particles to be equal to their SPH radii, the power spectrum of gas follows linear theory on large scales. On the other hand, gas looses power at high-redshift on small scales due to the relatively large smoothing lengths. That produces a suppression of power on small scales with respect to N-body simulations, that is also translated into a lower abundance of halos. While the problem of the lack of gas power on small scales can be alleviated by setting the smoothing lengths of the gas particles to their standard value, this procedure induces that the gas will not follow linear theory on large-scales due to its numerical gravitational coupling with CDM on small scales. 

Although both implementations have their pros and cons, our results indicate that the relative neutrino effects, are not largely affect ($\sim1\%$) by baryons down to $k=0.5~h{\rm Mpc}^{-1}$. Therefore, we confirm that neutrino effects can be factorized out and its magnitude can be determine through N-body simulations as long as calculations do not require sub-percent accuracy.\\

\end{appendix}

\bibliography{references}{}
\bibliographystyle{hapj}

\end{document}